\documentclass[]{aa}
\usepackage{todonotes}
\usepackage{graphicx}
\usepackage{natbib}
\usepackage[varg]{txfonts}
\usepackage{color}
\usepackage{longtable}
\usepackage{lscape}
\usepackage{url}
\usepackage{supertabular}

\newcommand\kms{\ensuremath{\mbox{km}\,\mbox{s}^{-1}}}
\newcommand\Teff{\ensuremath{T_\mathrm{eff}}}
\newcommand\logg{\ensuremath{\log g}}
\newcommand\vsini{\ensuremath{v\sin i}}

\newcommand\met{\ensuremath{\left[\dfrac{Fe}{H}\right]}}

\begin{document}

\title{A new method for the inversion of atmospheric parameters of A/Am stars.
\thanks{Based on data retrieved from the Polarbase, SOPHIE, and ELODIE archives}\fnmsep
}
\author{M. Gebran\inst{1} \and
W. Farah\inst{1}  \and
F. Paletou\inst{2,3}    \and
R. Monier\inst{4,5}    \and
V. Watson\inst{2,3}
}

\institute{
Department of Physics and Astronomy, Notre Dame University-Louaize, PO Box 72, Zouk Mika\"el, Lebanon
\and
Universit\'e de Toulouse, UPS-Observatoire Midi-Pyr\'en\'ees, IRAP, 
Toulouse, France
\and
CNRS, Institut de Recherche en Astrophysique et Plan\'etologie, 14 av. 
E. Belin, F--31400 Toulouse, France
\and
LESIA, UMR 8109, Observatoire de Paris, place J. Janssen, Meudon
\and
Laboratoire Lagange, Universit\'e de Nice Sophia, Parc Valrose,            06100 Nice, France
}
 
\date{Received / Accepted}

\titlerunning{Atmospheric parameters of A stars using PCA}
\authorrunning{M. Gebran et al.}

\abstract {We present an automated procedure that derives simultaneously the effective temperature \Teff, the surface gravity \logg, the metallicity [Fe/H], and the equatorial projected rotational velocity \vsini\ for "normal" A and Am stars. The procedure is based on the principal component analysis inversion method of \cite{S4n}.} {A sample of 322 high resolution spectra of F0-B9 stars, retrieved from the Polarbase, SOPHIE, and ELODIE databases, were used to test this technique with real data. We have selected the spectral region from 4400-5000\AA\ as it contains many metallic lines and the Balmer H$\beta$ line.}
 {Using 3 datasets at resolving powers of R=42\,000, 65\,000 and 76\,000, about $\sim 6.6\times 10^6$ synthetic spectra were calculated to build a large learning database. The Online Power Iteration algorithm was applied to these learning datasets to estimate the principal components (PC). The projection of spectra onto the few PCs offered an efficient comparison metric in a low dimensional space. The spectra of the well known A0- and A1-type stars, Vega and Sirius A, were used as control spectra in the three databases. Spectra of other well known A-type stars were also employed in order to characterize the accuracy of the inversion technique.} {All observational spectra were inverted and atmospheric parameters derived.  After removal of a few outliers, the PCA-inversion method appears to be very efficient in determining \Teff, [Fe/H], and \vsini\ for A/Am stars. The derived parameters agree very well with previous determinations. Using a statistical approach, deviations of around 150 K, 0.35 dex, 0.15 dex, and 2 km/s were found for \Teff, \logg, [Fe/H], and \vsini\ with respect to literature values for A-type stars.} {The PCA-inversion proves to be a very fast, practical, and reliable tool for estimating stellar parameters of FGK and A stars, and deriving effective temperatures of M stars.}
 \keywords{Stars: fundamental parameters -- Stars: early-type -- Methods: numerical}
 \maketitle 

%
%



\section{Introduction}
\label{intro}

Sky and ground-based surveys can produce hundreds of terabytes (TB) up to 100 (or more) petabytes (PB) both in the image data archive and in the object catalogs \citep{borne13}. For instance, the ESA space mission Gaia \citep{perryman01}, launched in December 2013, is expected to produce a final archive of about 1 PB ($10^{15}$ Bytes)  through 5 years of exploration. The Large Synoptic Survey Telescope (LSST) project will conduct a 10-year survey of the sky that will deliver about 15 terabytes (TB) of raw data per night \citep{kantor07}. The total amount of data collected over the ten years of operation will be 60 PB, and processing these data will produce a 15 PB catalog database. The need for methods that can handle and analyze this large amount of information becomes evident. 

The Principal Component Analysis (PCA) has been used to invert stellar spectra in order to determine stellar fundamental parameters of stars \citep{bailer, fiorentin, S4n, dms}. This technique has proven to be a fast and reliable tool to invert observed high-resolution spectra. Unlike classical techniques such as least-square fitting, PCA is based on a dimensionality reduction for the fast search of the nearest neighbour of an observed spectrum in a learning database. This technique can become obviously advantageous when dealing with large number of extended spectra collected at various spectral resolutions during sky surveys such as the Sloan Digital Sky Survey (SDSS; \citealt{sdss}), the RAdial Velocity Experiment (RAVE; \citealt{rave1,rave2}), and more recently the Gaia ESO Survey (GES; \citealt{ges}).

A  description of the PCA-based inversion technique for stellar parameters that we shall use can be found in \cite{S4n}, where the authors applied this technique to derive fundamental parameters from high-resolution spectra of FGK stars. \cite{dms} have shown that the range of application of the principal component analysis-based inversion method can be extended to dwarf M stars in order to derive effective temperatures.  

 In the present work, we extend the work of \cite{S4n} to A-type stars whose effective temperatures range between 7\,000 and 10\,000 K. A large variety of physical processes are at play in the envelopes of A/Am stars. Chemical peculiarity and pulsation are  observed among main-sequence A and F stars. Chemical peculiarity is a signature of the occurrence of transport processes competing with radiative diffusion \citep{zahn05,richer00,richard01,richard02,vick10}. Chemical peculiarity exist in different ways in A-type stars (Am, Ap, $\lambda$ Bootis stars$\ldots$). These peculiarities set constraints on self-consistent evolutionary models of these objects including various particle transport processes \citep{gebran08a,gebran08b,gebran10}. The stellar atmospheric parameters are the prerequisites to any detailed abundance analysis, and for that reason, these parameters should be derived with good accuracy. 

To do so, we have calculated around 6.6$\times$10$^6$ synthetic spectra and used them as three learning databases. These synthetic spectra were compared to a collection of observed spectra in order to find the best global fit between the two sets. The A-type stars spectra that we analyzed in the present work were selected from the PolarBase\footnote{\url{http://polarbase.irap.omp.eu}} \citep{polar} stellar library at two different resolutions ($R\sim$ 65\,000 and $R\sim$ 76\,000), and from the SOPHIE\footnote{\url{http://atlas.obs-hp.fr/sophie/}} and ELODIE\footnote{\url{http://atlas.obs-hp.fr/elodie/}} archives \citep{moultaka} ($R\sim$ 75\,000 and 42\,000, respectively). We have restricted our study to the wavelength range 4400-5000 \AA. In this region, free from telluric lines, we can find many metallic lines and mostly many iron transitions (Fe\,{\sc i} and Fe\,{\sc ii}) with accurate atomic data. The Balmer line H$\beta$ around $\sim$4860 \AA\ is also present in this spectral range. As mentioned by \cite{smalley2005}, the advantage of the global spectral fitting technique is that it can be automated for a very large number of stellar observations. It can also be used for low resolution spectra  or spectra of fast rotators where line blending is important (\vsini\ can go up to 300 \kms\ for A-type stars). Dealing with large multi-dimensional grid of synthetic spectra, the combination of the metallic and Balmer lines is essential for the derivation of the effective temperature \Teff, surface gravity \logg, metallicity [Fe/H], and projected equatorial rotational velocity \vsini\ \citep{smalley2005}. 

Due to their sensitivity to effective temperature and surface gravity, hydrogen line profiles should be included in all automatic procedures based on spectral fitting to obtain realistic atmospheric parameters for stars hotter than 5\,500 K \citep{ryabchikova2015}. The Balmer lines in B to F stars are in general well matched by LTE computations, and they are practically identical to the non-LTE predictions \citep{przybilla2004}. The region around H$\beta$ was chosen because it harbors several iron lines having accurate atomic parameters. The Balmer lines provide an excellent diagnostic of \Teff\ for stars cooler than $\sim$8\,000 K \citep{gray92,heiter2002}. For stars hotter than 8\,000 K, Balmer line profiles, in particular their wings, are sensitive to both temperature and gravity \citep{gray92,smalley2005}. Metal line diagnostics are also a good indicators of \Teff, \logg, and [Fe/H] as the best fit synthetic spectrum will correspond to the one that yields the same abundance for different ionization stages of the same element (Ionization Balance) and in the same abundance for all excitation potentials of the same element in a given ionization stage (Excitation Potential). On the other hand, lines such as the Mg\,{\sc ii} triplet around 4481 \AA\ and the Fe\,{\sc ii} lines between 4500 and 4530 \AA, are excellent indicators of \vsini\ (see for example \citealt{royer14}). For all the above mentioned reasons, we have applied the global spectral fitting technique to the whole spectral range from 4400 and 5000 \AA. The derived parameters are obviously model dependent, but the models that we are using (Sec.~\ref{learning}) are well suited for describing the atmospheres and the spectra of "normal" A and Am stars.

The selection of the observed spectra and the construction of the learning databases are detailed in Sec.~\ref{sect_sample}. The Principal Component Analysis and the derivation of the eigenvectors and the coefficients are recalled in Sec.~\ref{procedure}. Radial velocity correction and re-normalization of the spectra are discussed in Sec.~ \ref{RV}. The results of the inversion are presented in Sec.~\ref{inversion}. A few outliers have been analyzed in details in Sec.~\ref{outliers}. Discussion and conclusion are gathered in Sec.~\ref{conclusion}.

\section{Observations and learning database}
\label{sect_sample}

\subsection{Target selection}
\label{selection}
The PolarBase contains reduced stellar spectra collected with the NARVAL and ESPaDOnS high-resolution spectropolarimeters.  These two spectropolarimeters are mounted on the 2 m aperture \emph{T\'elescope Bernard Lyot} (TBL) in France and on the 3.6 m aperture CFHT telescope in Hawaii, respectively. Both instruments have a spectral resolving power of 65\,000 in polarimetric mode and 76\,000 when used for classical spectroscopy. 
As a first step, we have selected all high resolution \emph{Echelle} spectra of stars whose spectral type ranges from F0 up to B9 from the database at two different resolutions and in spectroscopy mode only. Fifty spectra were thus retrieved.

We have also retrieved A-type stars spectra from the ELODIE and SOPHIE archives. These archives are on-line databases of high-resolution \emph{Echelle} spectra and radial velocities.
ELODIE is a fiber-fed cross-dispersed \emph{Echelle} spectrograph that was attached to the 1.93-m telescope at OHP \citep{baranne96}. It
recorded in a single exposure a spectrum extending from 3850 \AA\
to 6811 \AA\ at a resolving power of about 42\,000 on a relatively
small CCD (1024 $\times$ 1024). SOPHIE is the \emph{Echelle} spectrograph mounted on the 1.93-m telescope at the Observatoire de Haute-provence (OHP) since September 2006. SOPHIE spectra stretch from 3870 to 6940 \AA\ in 39 orders with two different spectral resolutions: the high resolution mode HR (R$\sim$75\,000) and the high efficiency mode HE (R=39\,000). All the observations that we used in this work were observed in the HR mode at the same resolution of NARVAL spectra. Spectra of 279 stars, with signal to noise ratio larger than 100, were downloaded from the SOPHIE and ELODIE archives.
Overall, we have selected 322 stars for our study. The identifications of the selected stars are collected in Tab.~\ref{obs}. These data correspond to PolarBase, ELODIE, and SOPHIE spectra, respectively. Inverted effective temperatures, surface gravities, projected equatorial rotational velocities, metallicities, and radial velocities are collected in this table (see Sec.~\ref{inversion}).

\subsection{Learning database}
\label{learning}
The learning database was constructed using synthetic spectra as done in \cite{dms}. Three sets of grids were used, one at a resolution of 65\,000, one at 76\,000, and one at 42\,000, corresponding to the resolution of Polarbase, ELODIE and SOPHIE spectra respectively. These 3 grids are identical in terms of the parameters (\Teff, \logg, [Fe/H], \vsini, $\xi_t$), only the resolution differs. Line-blanketed ATLAS9 model atmospheres \citep{Kurucz} were calculated for the purpose of this work. ATLAS9 models are LTE plane parallel and assume radiative and hydrostatic equilibrium. This ATLAS9 version uses the Opacity Distribution Function (ODF) of \cite{castelli03}. We have included convection in the atmospheres of stars cooler than 8500 K. Convection was treated using a mixing length parameter of 0.5 for 7000 K$\leq$\Teff$\leq$ 8500 K, and 1.25 for \Teff$\leq$7000 K following Smalley's (2004) prescriptions.

The grid of synthetic spectra was computed using SYNSPEC48 \citep{Hubeny}. The effective temperatures of the model atmospheres range from 6\,800 and 11\,000 K with a step of 100 K, and logarithm of surface gravities between 2.0 and 5.0 with a step of 0.1 dex, respectively. The projected equatorial rotational velocities \vsini\ were varied from 0 up to 300 \kms\ with a non-constant step (2, 5, and 10 \kms), metallicity was scaled from -2.0 dex up to +2.0 dex with respect to the \cite{Grevesse98} solar value with a step of 0.1 dex. The synthetic spectra were computed from 4400 \AA\ up to 5000 \AA\ with a wavelength step of 0.05 \AA. This range contains many moderate and weak metallic lines in different ionization stage. These weak metallic lines (indicators for \vsini\ and [Fe/H]) and the Balmer line (indicator for \Teff\ and \logg) are not sensitive to small changes of the microturbulent velocity. For that reason, we have fixed the microturbulent velocity ($\xi_t$) to 2.0 \kms in all models. This value corresponds to the average microturbulent velocity in A-type stars \citep{gebranxi}.
The linelist used in the synthetic spectra calculation was constructed from Kurucz gfhyperall.dat\footnote{\url{http://kurucz.harvard.edu}} and modified with more recent and accurate atomic data retrieved from the VALD\footnote{\url{http://www.astro.uu.se/~vald/php/vald.php}} and the NIST\footnote{\url{http://physics.nist.gov}} databases. At each resolution, we ended up with $\sim$ 2.2 million spectra as a learning database. 
 
\section{Principal Component Analysis}
\label{procedure}
Our usage of the PCA for the inversion process is 
influenced by the work of \cite{rees00} and \cite{S4n,dms}. As mentioned in Sec.~\ref{intro}, the PCA is a reduction of dimensionality technique that retains as much as possible the variation present in the data set \citep{Jolliffe}. It searches for basis vectors that represent most of the variance in a given database.
The reduction of dimensionality allowed by the PCA
is directly used to build a specific metric from which a nearest-neighbour(s) search is made between an observed data set (observed spectra in the present work) and the content of a learning database (synthetic spectra in the present work). The learning database could be constructed using a set of observed spectra with well known parameters \citep{S4n}. The PCA becomes mainly advantageous when dealing with high- and low-resolution spectra that cover a very broad bandwidth as the reduction of dimensionality will allow a very fast processing of the data. In what follows, we describe the characteristics of this technique in the context of our study.

\subsection{Nearest neighbour search}
 Representing the learning database in a matrix $\textbf{\textit{S}}$ of size $N_{\mathrm{spectra}}=2.2\times10^6$ by $N_{\lambda}=12000$, $\bar{S}$ will be the average of $\textbf{\textit{S}}$ along the $N_{\mathrm{spectra}}$-axis. Then, we derive the eigenvectors $\textbf{e}_k(\lambda)$ of the variance-covariance matrix $\textbf{\textit{C}}$ defined as 

\begin{equation}
\textbf{\textit{C}}=(\textbf{\textit{S}}-\bar{\textit{S}})^\mathrm{T}\cdot(\textbf{\textit{S}}-\bar{\textit{S}})\, 
\end{equation}

and having a dimension of $N_{\lambda}\times N_{\lambda}$. These eigenvectors of the symmetric $\textbf{\textit{C}}$ matrix are then sorted in decreasing eigenvalues magnitude; they are more usually referred to as ``principal components''. Each spectrum of the dataset $\textbf{\textit{S}}$ is represented by a small number of coefficients $p_{jk}$ defined as 
\begin{equation}
p_{jk}=(S_j-\bar{S})\cdot \textbf{\textit{e}}_k
\end{equation}
 The main point of reducing the dimensionality is the selection of $k_{max}\ll N_{\lambda}$. We show in Sec.~\ref{power} that $k_{max}=12$ is a reasonable choice.\\

 To follow the same notation as in \cite{S4n}, we denote the observed spectrum by $O(\lambda)$ having the same wavelength range and sampling as the learning dataset spectra. At this stage, the observations should be corrected for radial velocity (details in Sec.~\ref{RV}). We compute the set of projection coefficients $\varrho_k$ defined as
\begin{equation}
\varrho_k=(O-\bar{S})\cdot \textbf{\textit{e}}_k
\end{equation}
 The nearest neighbour is then found by minimizing a $\chi^2$ in the low dimensional space of the coefficients 
 \begin{equation}
 d^{(O)}_j=\Sigma^{k_{max}}_{k=1}(\varrho_k - p_{jk})^2,
 \end{equation}
 where $j$ covers the number of synthetic spectra.  The parameters of the synthetic spectrum having the minimum $d$ will be considered for the inversion.

\subsection{Power Iteration method}
\label{power}

The derivation of the leading eigenvectors of $\textbf{\textit{C}}$ is the most critical step when performing PCA, and it becomes increasingly time consuming as the size of learning dataset increases. Furthermore, techniques such as singular value decomposition (SVD) that were used in previous studies, become inapplicable in this case due to computer memory shortage. In our particular case, at each of the three resolutions, the files containing the matrix $\textbf{\textit{S}}$ are $\approx$ 240 GB in size. To bypass this problem, we have used an algorithm that is applicable independently of the size of the dataset. We have implemented the \textit{Online} Power Iteration algorithm (also called \textit{Von Mises} iteration, \citealt{demel97}) to estimate the $k_{max}$ principal components of the dataset $\textbf{\textit{S}}$. This algorithm reads and operates on synthetic spectra of the learning dataset one by one, avoiding loading all the data at once. The procedure, moreover, does not need any typical matrix factorization such as SVD, and is efficient when applied to large datasets \citep{roweis}. The procedure consists of the following:

First we find the eigenvector $e_1$ with a maximal eigenvalue $\lambda_1$ (i.e. the first principal component), by iterating 
\begin{eqnarray}
x_{n+1}=\dfrac{\textbf{\textit{C}}x_n}{||\textbf{\textit{C}}x_n||},
\label{pi}
\end{eqnarray}
with an initial random non-zero vector $x_0$. Practically $\textbf{\textit{C}}x_n$ cannot be computed using standard technique as one should load the full dataset $\textbf{\textit{S}}$ at once to calculate $\textbf{\textit{C}}$.
To overcome this problem, we evaluate "$\textbf{\textit{C}}x_n$" for each iteration on $x_n$ while operating on each spectrum at a time by performing: 
\begin{eqnarray}
\textbf{\textit{C}}x_n &=& (\textbf{\textit{S}}-\bar{S})^T (\textbf{\textit{S}}-\bar{S})\cdot x_n \nonumber \\
&=& \sum_{i=1}^{N_{spectra}}((S_i-\bar{S})\cdot x_n)\cdot (S_i-\bar{S})
\label{pi2}
\end{eqnarray}

\noindent where $S_i$ is the $i^{th}$ spectrum in the dataset $\textbf{\textit{S}}$. The $x_n$ iterations converge to the eigenvector $e_1$ that corresponds to the largest eigenvalue. 

  Once we find a good approximation for $e_1$, we perform the Gram-Schmidt process in order to remove the contribution of $e_1$ from $\textbf{\textit{S}}$ by doing:
\begin{equation}
\textbf{\textit{S}}_1= \textbf{\textit{S}}-[(\textbf{\textit{S}}-\bar{S})\cdot e_1]e_1
\end{equation} Again, the above should be done iteratively over each spectrum. Next, to compute $e_2$, the power method (Eq.~\ref{pi}) is performed on \begin{equation}
\textbf{\textit{C}}_1=(\textbf{\textit{S}}_1-\bar{S})^T\cdot (\textbf{\textit{S}}_1-\bar{S})
\end{equation}
 Once $e_2$ is estimated, the Gram-Schmidt process is applied again to yield: 
 \begin{equation}
  \textbf{\textit{S}}_2=\textbf{\textit{S}}_1 - [(\textbf{\textit{S}}_1-\bar{S})\cdot e_2] e_2
  \end{equation}
And so on until we compute all the needed eigenvectors. To determine the optimal number of iterations needed for the derivation of the eigenvectors, we have constructed several smaller databases and derived their respective eigenvectors using the SVD technique and the Power Iteration. For each database, the derived eigenvectors from both techniques were compared. Using Power Iteration, 25 iterations were required to reach a relative difference between eigenvectors smaller than 0.1\%. The use of the Power Iteration in this present work allowed the increase of the number of spectra in the learning database by a factor of 10 (200 000 in case of the SVD to 2 200 000 in case of Power Iteration).\\

Only the first 12 eigenvectors were considered and that was justified by the calculation of the reconstruction error of the spectra defined as
\begin{equation}
E(k_{max})=\left< \left( \dfrac{\vert\bar{S}+\Sigma_{k=1}^{k_{max}} p_{jk}\textbf{\textit{e}}_k - S_j\vert}{S_j}    \right)  \right>
\end{equation}
Using only 12 eigenvectors, the average reconstruction error is found to be smaller than 1\%, a result similar to that derived in \cite{S4n}.

\section{Preparation of the observed spectra}
\label{RV}
In order to adjust a synthetic spectrum to an observed one, lines of both spectra should be aligned, as accurately as possible, on the same wavelength scale and both fluxes should be at the same level. The later is achieved by rectifying the observed and synthetic spectra to their local continua.

\subsection{Radial velocity correction}
A wavelength shift between observations and synthetic spectra could affect drastically the derived parameters of the first neighbour. \cite{S4n} showed that the radial velocity $v_r$ should be known to an accuracy of $c/4R$, where $c$ is the speed of light in vacuum. This requirement translates into a velocity accuracy of $\sim$1 \kms\ needed at a spectral resolution of 76\,000. We have corrected all the spectra from radial velocity (given in the last column of Tab.~\ref{obs}) before the inversion procedure. The radial velocities were determined using the classical cross-correlation technique originally described by \cite{ccf}. The observed spectra were cross-correlated using a synthetic template, in the same wavelength range, corresponding to the parameters $\Teff=8500$\,K, $\logg=4.0$\,dex, [Fe/H]=0, and $v_e\sin i=0$\,\kms. Many of the Fe\,{\sc ii}, Cr\,{\sc ii}, Ti\,{\sc ii} and Mg\,{\sc ii} atomic lines are mainly present in all sub-types of A-type stars (from A0 to A9). The wavelength of the Fe\,{\sc ii} lines in the 4500-4550 \AA\ region are accurately measured to within $\sim$ 0.01 \AA\ according to the NIST database. For that reason, the selected template corresponds to a typical A5V type star.

\subsection{Renormalization of the flux}

\begin{figure}[!h]
\centering
\resizebox{\hsize}{!}{\includegraphics{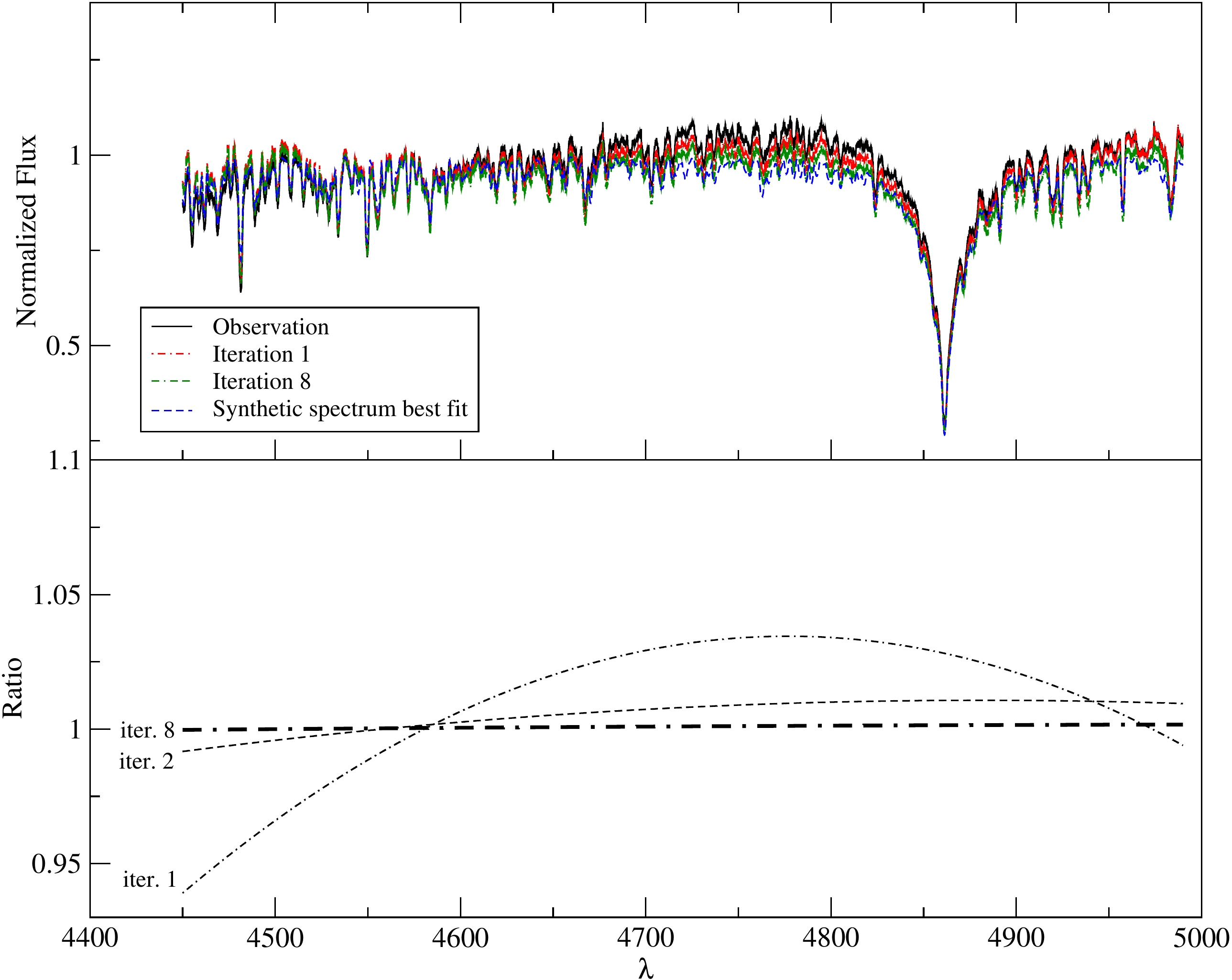}}
\caption{Upper panel: Continuum correction using 8 iterations for the inversion of HD97230. Only iterations 1 and 8 are shown for the sake of clarity, The initial observation is in full line, the 1$^{st}$ iteration is in dashed-dot style, the 8$^{th}$ iteration is in double dashed-dot style, and the final best fit synthetic spectrum is in dashed style. Bottom Panel: Ratios between the re-normalized spectrum of the observation and the first neighbour best fit synthetic spectrum for iterations 1, 2, and 8.}
\vskip 0.05cm
\label{HD97230}
\end{figure}

On the other hand, we also noticed that the flux normalization procedure is critical in this work and that an improper normalization could affect drastically the derived parameters. In order to correct for this effect, we have proceeded using \cite{Gazzano}'s procedure. Each originally normalized observed spectrum was divided by the synthetic spectrum calculated with the first inverted atmospheric parameters (\Teff, \logg, [Fe/H], and \vsini). The residuals were then cleaned from lines with a sigma clipping, rejecting points above or below 1 $\sigma$. The remaining points were fitted with a second degree polynomial. The observation was then divided by this polynomial in order to obtain a new normalized spectrum. This new spectrum was inverted again in order to find a new set of parameters. This procedure was repeated until the inverted parameters remained unchanged from one iteration to the next. The number of iterations depends on the original shape of the normalized observed spectrum. On average, 10 iterations were sufficient in order to reach convergence. 
In Fig.~\ref{HD97230}, we show the application of Gazzano et al. (2010) procedure to the inversion of the spectrum of \object{HD~97230}. Only 8 iterations were required in this example.

\section{A stars spectra inversions}
\label{inversion}

\subsection{Vega and Sirius A}
\begin{figure}[!htp]
 \centering
\includegraphics[width=\hsize]{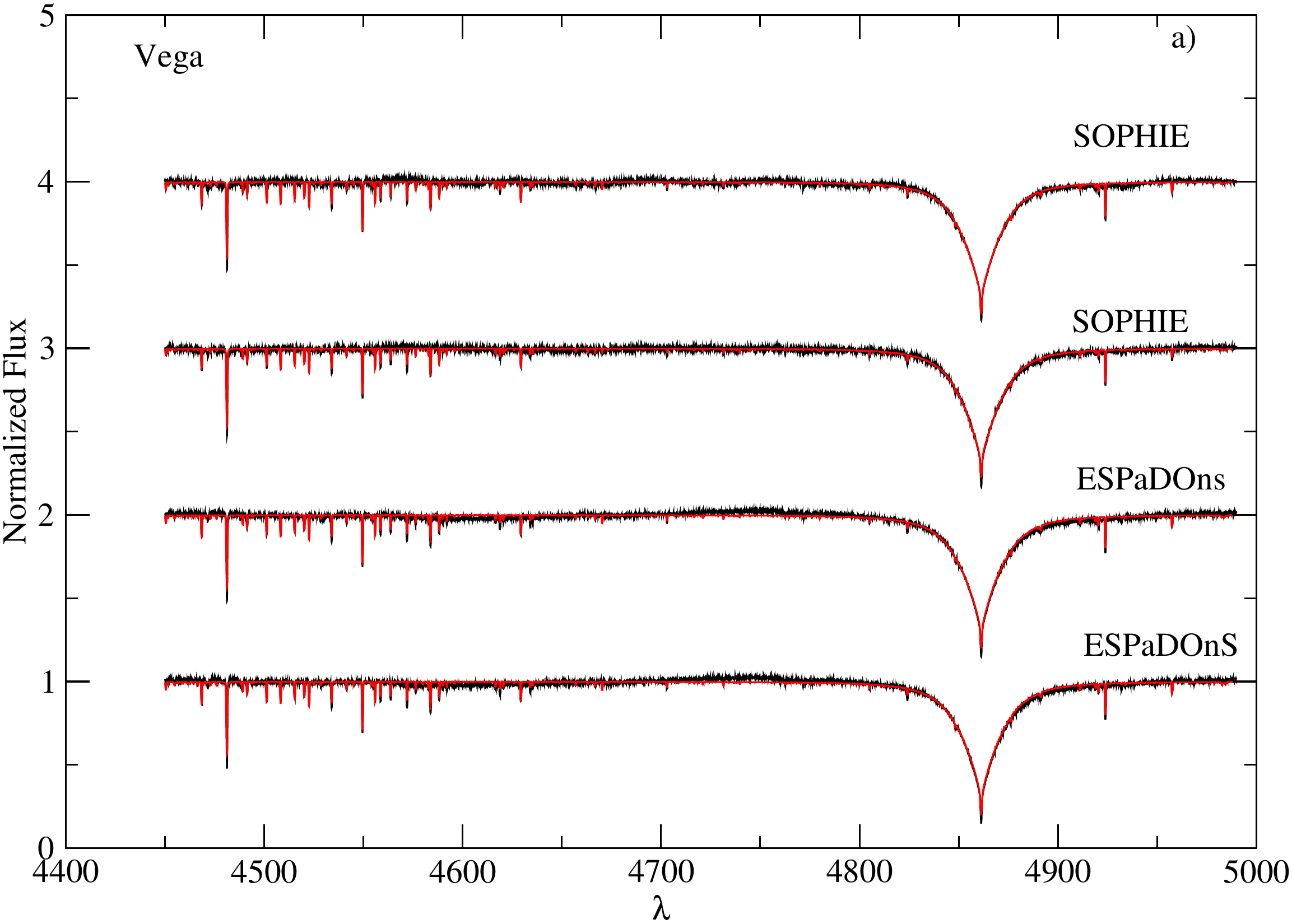}\\
\includegraphics[width=\hsize]{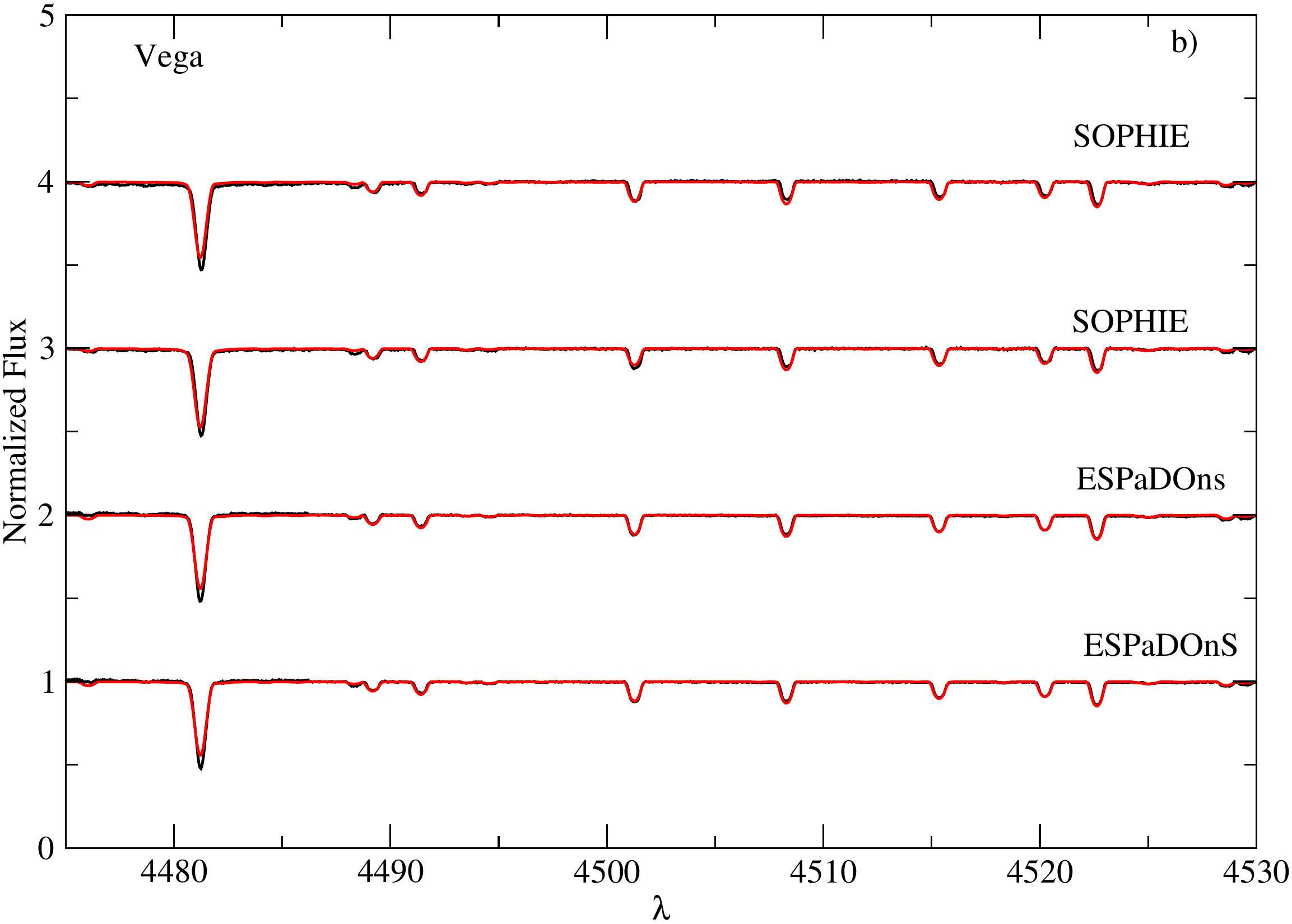}
\caption{First neighbour fit of synthetic spectra to the observed ones for the 4 spectra of Vega, observed at different epochs, retrieved from PolarBase and SOPHIE databases. Observed spectra are in black and the synthetic ones are in red. Part a) represents the overall spectrum whereas part b) displays the region sensitive to [Fe/H] and \vsini. The wavelengths are in \AA. Spectra have been offset by arbitrary amounts.}
\label{figvega}
\end{figure}

The first tests were carried out using the two well studied A-type stars, Vega (HD~172167, A0V) and Sirius A (HD~48915, A1m). The spectra of Vega were retrieved from the PolarBase and SOPHIE archive. The ESPaDOnS spectrum was inverted using the learning database at a resolution of 65\,000 whereas the SOPHIE spectrum was inverted using the database at a resolution of 76\,000. Both inversions yielded the same results for all parameters. We found for $\lbrace$\Teff, \logg, [Fe/H], \vsini$\rbrace$ values of $\lbrace$9\,500 K, 4.0 dex, -0.5 dex, 25 \kms$\rbrace$ whereas the medians of the parameters found by querying Vizier\footnote{\url{http://vizier.u-strasbg.fr/viz-bin/VizieR}} are $\lbrace$9\,520 K, 3.98 dex, -0.53 dex, 23 \kms$\rbrace$. The inverted parameters of Vega agree very well with all previous determinations such as \cite{takeda08}, \cite{hill10}, and the fundamental effective temperature derived by \cite{code76} from the integrated flux and the angular diameter.
Figure~\ref{figvega}a displays the overall adjustment of 4 observed spectra of Vega with the first neighbour synthetic spectra. These spectra were observed at different epochs. The good agreement between the inverted metallicity and the projected equatorial rotational velocity of Vega with the catalogs ones can be seen in Fig.~\ref{figvega}b where we display the adjustment in the region sensitive to these two parameters.
The spectrum of Sirius A was retrieved from the ELODIE database. This spectrum was inverted using the learning database at a resolution of 42\,000. We found for $\lbrace$\Teff, \logg, [Fe/H], \vsini$\rbrace$ values of $\lbrace$9\,800 K, 4.2 dex, 0.2 dex, 18 \kms$\rbrace$ whereas the medians of the parameters retrieved from a Vizier query are $\lbrace$9\,870 K, 4.30 dex, 0.36 dex, 15 \kms $\rbrace$. This result is in very good agreement with previous findings such as \cite{hill93}, \cite{takeda09}, and \cite{landstreet11}.

Table~\ref{obs} displays the inverted parameters, the values that we retrieved from Vizier queries closest to our inverted ones, and the median of queried catalogs values. All literature data were collected from Vizier catalogs using the ASTROQUERY\footnote{\url{astroquery.readthedocs.org}} Python modules {\tt \citep{2014arXiv1408.7026P}}. We kept in this table the inverted parameters for all selected stars in spite of the fact that many of them may be variables or binaries. Using Simbad queries, we have listed, in the last column of Tab.~\ref{obs}, the peculiarity of the stars. Variables and binary systems were discarded in the characterization of the technique (Sec.~\ref{reference}).

\begin{figure*}[!t]
 \centering
\includegraphics[scale=0.42]{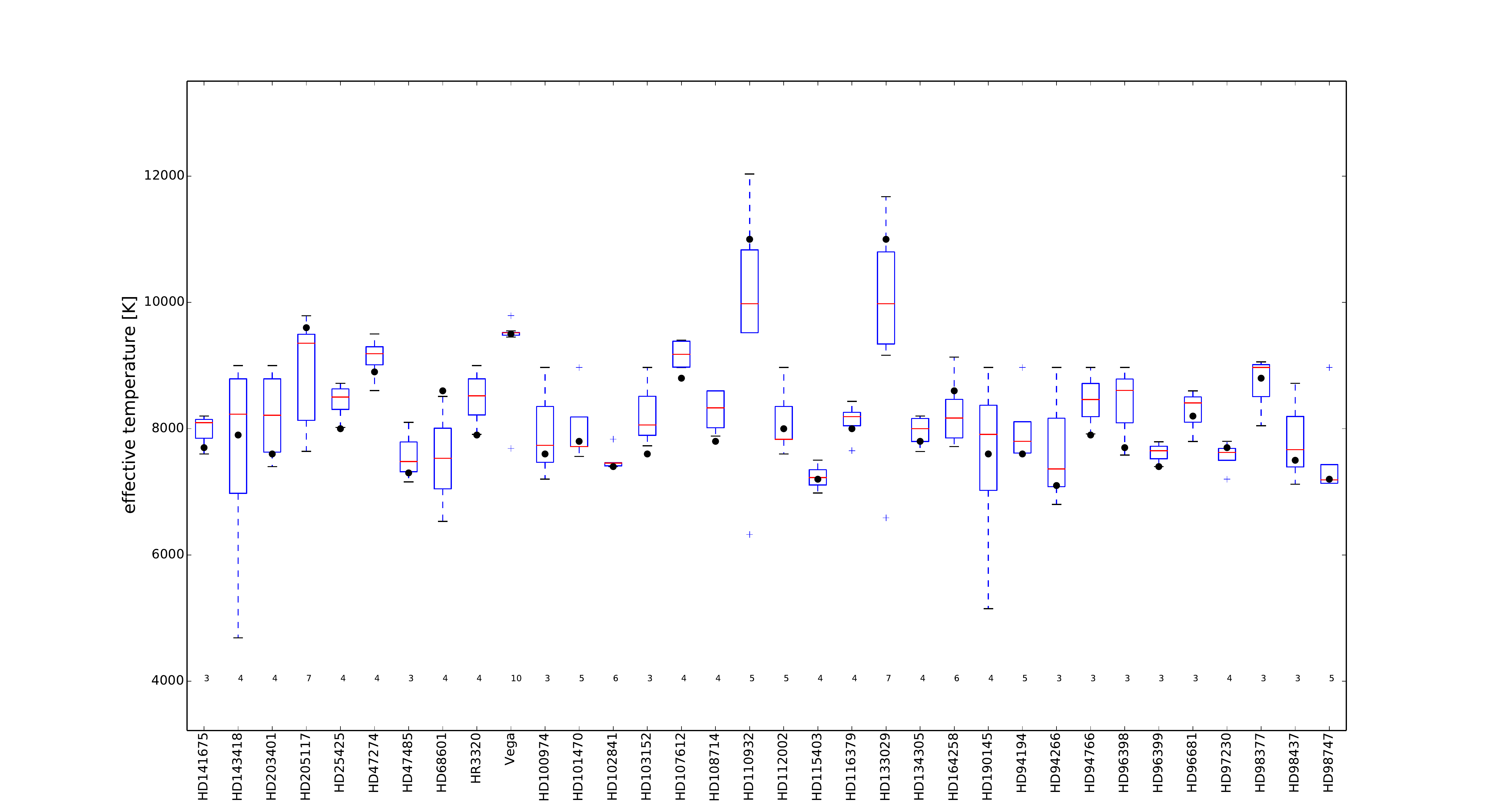}
\caption{Comparison between our estimate of effective temperatures ($\bullet$), and the values we got from available Vizier catalogs, for PolarBase spectra. The latter collections are represented as classical boxplots. Objects we studied are listed along the horizontal axis. The number of values found among all Vizier catalogs are presented on a horizontal line around T$_{\rm{eff}}\sim$ 4000 K. The horizontal bar inside each box indicates the median ($Q_2$ value) while each box extends from
first quartile, $Q_1$, to third quartile $Q_3$. Extreme values, still within a 1.5 times the interquartile range away from either $Q_1$ or $Q_3$, are connected to
the box with dashed lines. Outliers are denoted by a “+” symbol.}
\vskip 0.1cm
\label{Teff}
\end{figure*}

\begin{figure}[!htp]
 \centering
\includegraphics[width=\hsize]{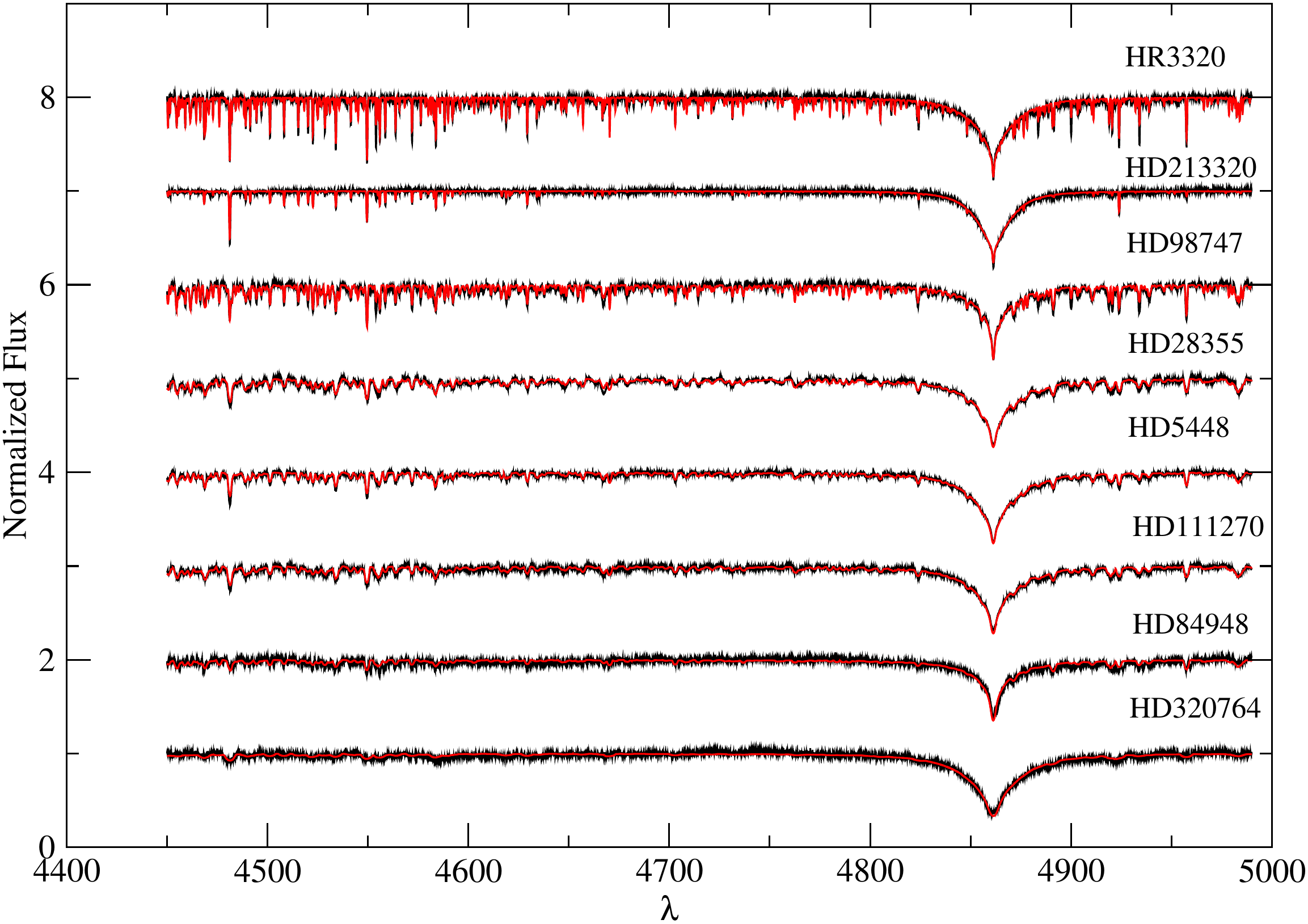}
\caption{First neighbour fit of synthetic spectra to the observed ones. The wavelengths are in \AA. Spectra have been shifted up for clarity reasons.}
\label{fit}
\end{figure}

The effective temperature is the only atmospheric parameter which can be retrieved from Vizier queries for all stars but for one exception. A small number of the selected stars do not have catalogs values for \logg, \vsini, and [Fe/H]. Figure~\ref{Teff} displays the inverted effective temperature for each spectrum retrieved from Polarbase as well as the range in the effective temperatures retrieved from the catalogs (box-plots) and the median. The number of values found in all catalogs retrieved in Vizier for each star also appears in this figure. We found that the inverted parameters agree well with those retrieved in the various catalogs. This agreement also applies to A-type stars spectra retrieved from ELODIE and SOPHIE archives (Tab.~\ref{obs}). However, we found several outliers which are discussed in detail in Sec.\ref{outliers}. These outliers are depicted with a symbol "*" in Tab.~\ref{obs}. 
Figure~\ref{fit} shows an example of an overall spectral fitting between the first neighbour synthetic spectrum and the observation for several A-type stars of our study.

\subsection{Evaluation of the results and Comparison with literature data}
\label{reference}
In order to evaluate the efficiency of our method, we have first removed from the analysis all peculiar stars such as Ae/Be, Ap, variable ($\delta$ Scuti, $\gamma$ Doradus variables$\ldots$) and binary stars (Eclipsing binaries, spectroscopic binaries$\ldots$) as the models that we are using do not take into account such peculiarities. ATLAS12 \citep{kurucz05} model atmospheres, dealing with Opacity Sampling (OS), can better reproduce the distribution of the thermodynamical quantities in the atmosphere of chemically peculiar stars. Peculiarities of Am stars are not strong enough to alter the temperature distribution in their atmospheres \citep{gebran08a,balona12}. For that reason, we are confident that ATLAS9 can be used for moderate chemical peculiarity.

We ended up with 182 classified "normal" A or Am stars. To assess an \emph{automated and objective} comparison with already published values one may find in Vizier@CDS, for instance, we found reasonable to adopt specific weights for the set of values we could retrieve for each object. These weights directly rely on the spread of catalogued values. For some of the stars, dispersions in the catalogued values were noticed, and were severe up to thousands of Kelvin for \Teff. As an example, HIP109119 has catalogued \Teff\ ranging between 5309 K and 8970 K. On the other hand, some stars had only one or no catalogued values at all for a particular parameter. This would definitely add a bias to the error calculation. For that reason we have used a statistical approach that gives less weight to stars having large spread in the catalogued values. For a star $i$ and a parameter $\psi$ ($\psi$ could be one of the following parameters: \Teff, \logg, [Fe/H], \vsini), the difference between the inverted and the median from the catalogued values of this parameter is weighted with the inverse of the interquartile range ($IQR$). $IQR$ is defined as the difference between the third and the first quartile of each set of values ($IQR=Q_3-Q_1$). The deviation for the parameter $\psi$ is derived following:

\begin{equation}
\Delta \psi_w = \dfrac{\Sigma_i w_i (\psi^{inv}_i - \psi^{med}_i)}{\Sigma w_i} \label{deviation}
\end{equation}

with $w_i$ derived for each star and each parameter as follow: 
\begin{equation}
w_i=\dfrac{1}{IQR_i}
\end{equation}

The associated standard deviation is found  according to the following equation:
\begin{equation}
\sigma^2_{w} = \dfrac{\Sigma_i w_i (\psi^{inv}_i - \psi^{med}_i)^2}{\Sigma w_i} \label{sigmadeviation}
\end{equation}

Stars with fewer than two catalogued value for a parameter $\psi$ were omitted in the calculation of $\Delta \psi_w$ because of the zero value of their $IQR$. To estimate the parameters spread in the catalogued values for the remaining stars, we have calculated the average $IQR$ as well as the standard deviation. We found that, on average, the interquartile values for \{\Teff, \logg, [Fe/H], \vsini\} are \{550 K, 0.25 dex, 0.17 dex, 10.7 \kms\} with a standard deviations of \{1200 K, 0.35 dex, 0.23 dex, 15.5 \kms\}. These numbers show that a large spread exists in the literature values, validating our approach in using a weighted mean. Applying Eq.~\ref{deviation}, the derived deviations on the 4 parameters and their corresponding standard deviations are summarized in columns 2 and 3 of Tab.~\ref{accuracy}. Columns 4 to 11 of Tab.~\ref{accuracy} display the absolute mean signed differences and the standard deviations obtained by comparing our inverted parameters to the median and closest ones from Vizier query for all stars (Columns 4 to 7) and for the 19 reference stars (Columns 8 to 11, see Sec.~\ref{19stars}).


\begin{table*}[!th]
\begin{center}
\begin{tabular}{|c|c|c|c|c|c|c|c|c|c|c|}
\hline

 &	$\Delta_w$		&$\sigma_{w}$&	$\Delta_m$ & $\sigma_{m}$ &$\Delta_c$ & $\sigma_{c}$ &$\Delta_{19m}$ & $\sigma_{19m} $  &$\Delta_{19c}$ & $\sigma_{19c}$ \\
 \hline
 \Teff\ (K)&	150  &500 &110&630& 13&220&35& 250& 8.5&96 \\ \hline
 \logg\ (dex)& 0.35&  0.30&0.40& 0.60 &0.40 &0.30 &0.18&  0.40&0.17&0.30    \\ \hline
 [Fe/H] (dex)& 0.15& 0.25&0.15& 0.25 & 0.12&0.20&0.09& 0.15& 0.08&0.13 \\ \hline
 \vsini\ (\kms)& 2.0 & 9.5&5.0& 15.0 &1.7&10.0 &4.0& 9.5 &2.0&5.5\\ \hline
 \end{tabular}
\caption{Average weighted deviations between our inverted parameters and the catalogued values ($\Delta_w$), with their respective standard deviation ($\sigma_{w}$). The $\Delta_m$, $\sigma_{m}$, $\Delta_c$, $\sigma_{c}$, display the unweighed average mean signed differences and their respective standard deviations between our inverted parameters and the median (m)/closest (c) of the catalogues values, for the 182 and the 19 reference ($\Delta_{19}$ and $\sigma_{19}$) stars.}\label{accuracy}
\end{center}
\end{table*}

\subsubsection*{Effective temperature \Teff}
\begin{center}
 \begin{figure*}[!tbh]
\includegraphics[scale=0.35]{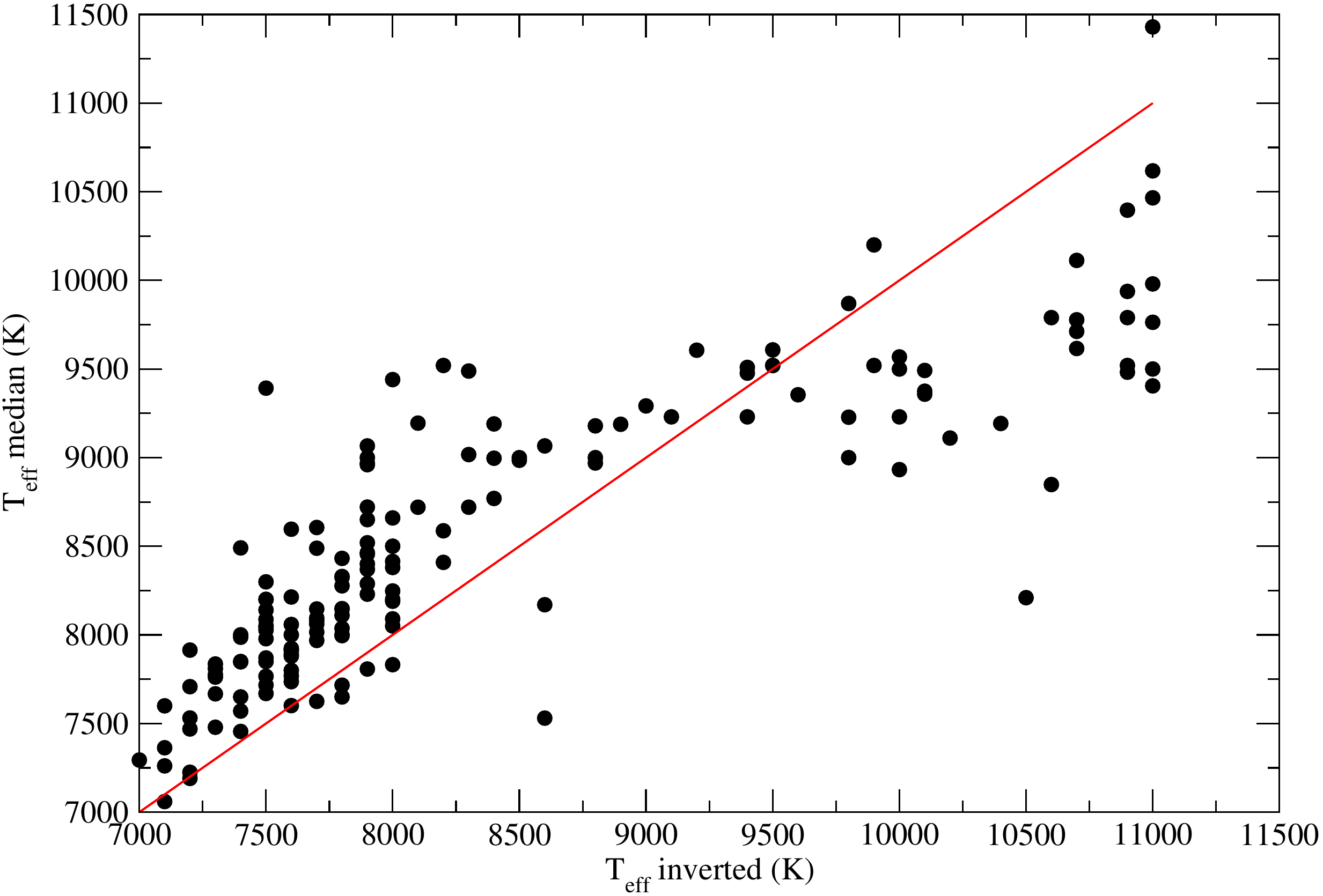}
\includegraphics[scale=0.35]{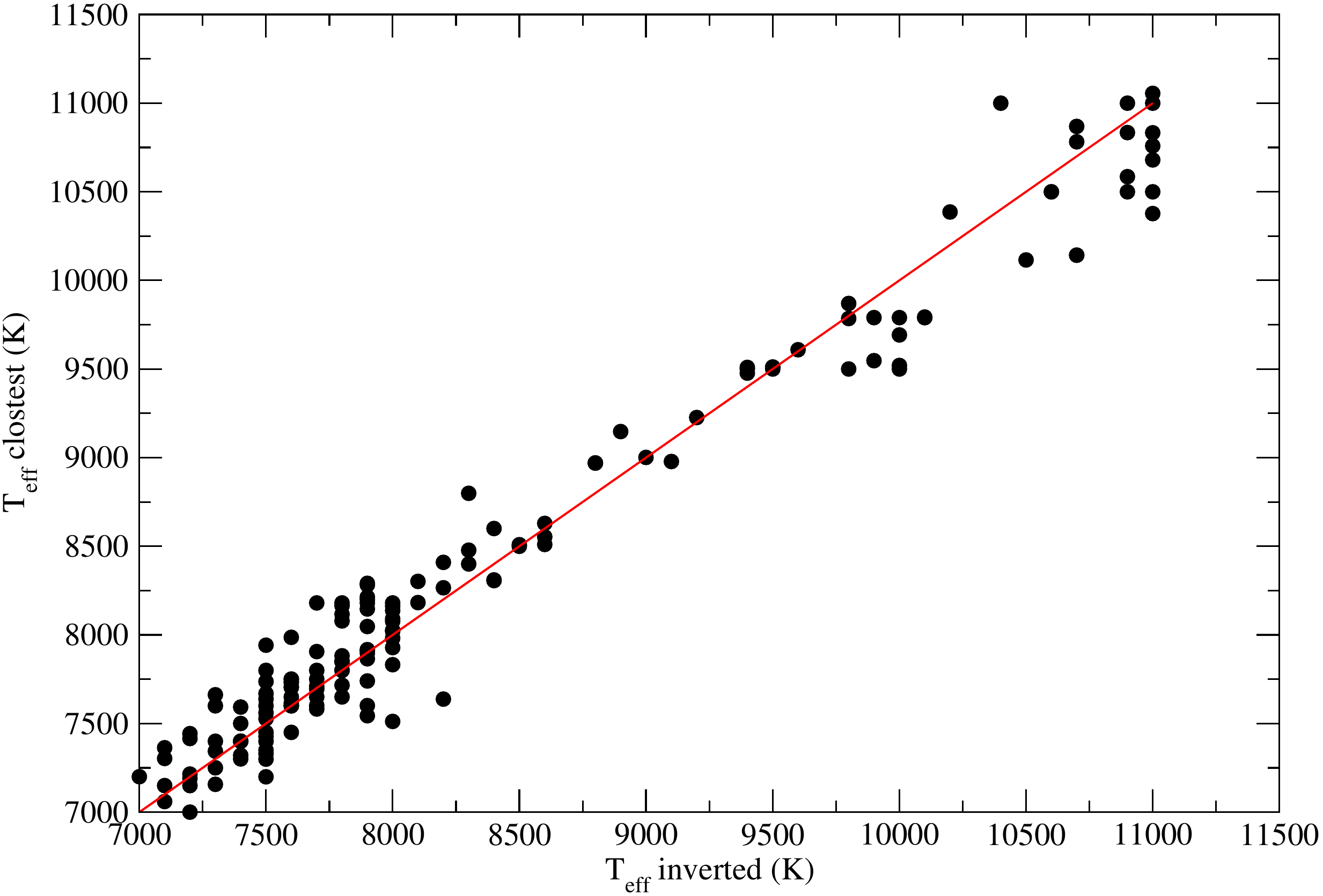}
\caption{Left panel: Comparison between our inverted effective temperatures and the median of the catalogued ones. Right panel: Comparison between our inverted  effective temperatures and the closest of the catalogued ones. }
\label{compar}
\end{figure*} 
\end{center}

Figure~\ref{compar} displays the comparison between our inverted effective temperature and the catalogued ones for the 182 stars. The left panel of Fig.~\ref{compar} represents the comparison with the median of the catalogued values whereas the right one displays the comparison with the closest. 
There is a systematic trend noticed in the left plot of Fig.~5. This is partly due to the large spread in the catalogued data. By comparing our inverted \Teff\ to the median value, we obtain an absolute mean signed difference of 110 K with a standard deviation of 630 K (Columns 4 and 5 of Tab.~\ref{accuracy}). All stars (except HD322676) have at least two values for catalogued \Teff. These catalogued effective temperatures are based on different bibliographical references using distinct tools/techniques for their determination. Using spectroscopy, one should expect some differences between the derived effective temperatures and the ones determined using photometric techniques. The global spectral fitting that includes the Balmer lines provides an excellent \Teff\ diagnostic for stars cooler than $\sim$8000 K because they do not depend on gravity \citep{gray92}. It follows that one part of the differences between inverted and catalogued values is justified for stars having \Teff$>$8000 K. For stars hotter than 8000 K, a degeneracy could exist between the \Teff\ and \logg. To overcome this degeneracy, we have selected a wavelength range containing many metallic lines as described in Sec.~\ref{intro}. When comparing our inverted effective temperatures with the closest catalogued values, the correlation becomes clearer (correlation coefficient of 0.98). The absolute mean signed difference and standard deviation are found to be, in that case, 13 K and 220 K, respectively (Columns 6 and 7 of Tab.~\ref{accuracy}). Taking into account the spread of catalogued data by applying Eq.~\ref{deviation}, we found that on average our effective temperatures deviate of about 150 K from the catalogued ones with a standard deviation of 500 K (Columns 2 and 3 of Tab.~\ref{accuracy}).

\subsubsection*{Surface gravity \logg}

As mentioned in Sec.~\ref{intro}, the H$\beta$ line is not sensitive to gravity for \Teff$\leq$8\,000 K, which applies to about 60\% of our sample. For that reason we have expected that the global spectral fitting technique could derive incorrect \logg\ values for these stars. The most reliable \logg\ values for A-type stars are determined using photometric techniques, more precisely, from the $uvby\beta$ calibration. The photometric values of \logg\ are not affected by metallicity \citep{smalley93}. By using a combination of spectroscopic and photometric techniques, the typical errors on the atmospheric parameters of a star is $\pm$100 K for \Teff\ and $\pm$0.2 dex for \logg\ \citep{smalley2005}. The 0.35 dex weighted average deviation and its corresponding 0.30 dex standard deviation found in the present work (Columns 2 and 3 of Tab.~\ref{accuracy}), by applying Eq.~\ref{deviation}, are justified by the use of only spectroscopic fitting technique and by the insensitivity of the H$\beta$ line to gravity for \Teff$\leq$8\,000 K. Comparing our inverted \logg\ to the median of the catalogued ones, we found an unweighed absolute mean signed difference of 0.40 dex with a standard deviation of about $\sim$0.60 dex (Columns 4 and 5 of Tab.~\ref{accuracy}).

\subsubsection*{Metallicity [Fe/H]}

Catalogued values for metallicities exist for a small number of stars and most of them have only one determination of [Fe/H]. Statistically, our comparison with catalogued values is not reliable in this case but the 0.15 dex that we derived using our statistical approach (Eq.~\ref{deviation}) is an average deviation for the derived metallicities. The standard deviation in that case is 0.25 dex (Eq.~\ref{sigmadeviation}). These values are identical to the ones derived using the unweighed differences (Columns 4 and 5 of Tab.~\ref{accuracy}). Comparing with the closest catalogues values, these numbers are found to be 0.12 dex and 0.20 dex, respectively (Columns 6 and 7 of Tab.~\ref{accuracy}).

\subsubsection*{Projected equatorial rotational velocity \vsini}

\begin{center}
 \begin{figure*}[!t]
\includegraphics[scale=0.365]{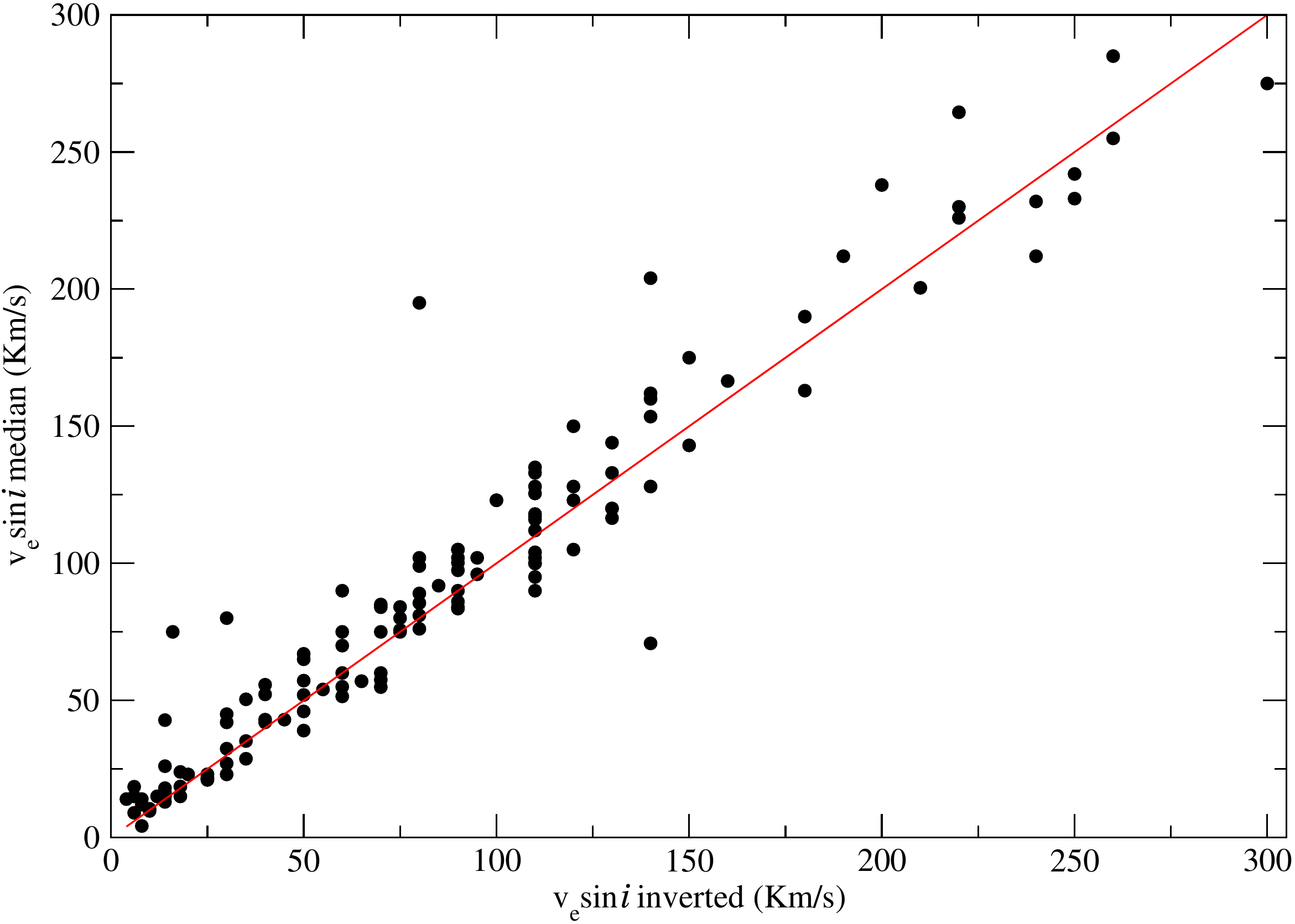}
\includegraphics[scale=0.365]{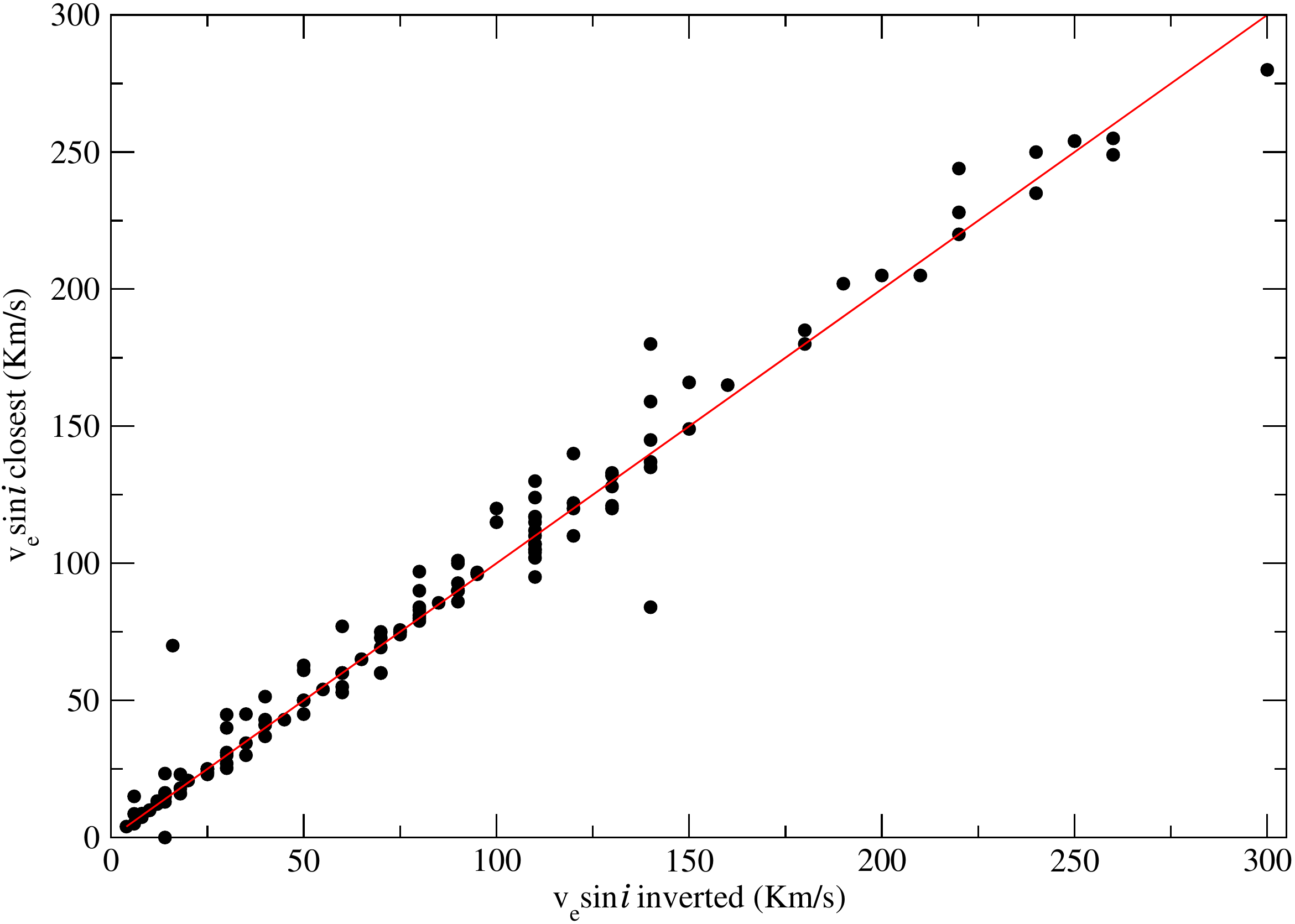}
\caption{Left panel: Comparison between our inverted projected equatorial rotational velocities and the median of the catalogued ones. Right panel: Comparison between our inverted projected equatorial rotational velocities and the closest of the catalogued ones.}
\label{compar-vsini}
\end{figure*} 
\end{center}

Most of the catalogued \vsini\ are based on the computation of the first zero of Fourier transform (FT) of \cite{royer02}. Our spectroscopic \vsini\ agree very well with previous findings as shown in Fig.~\ref{compar-vsini} where the comparison was done with the median and closest catalogued values. We found a correlation coefficient of 0.96 between our inverted values and the median ones from Vizier catalogs with a regression coefficient (slope) of 0.98$\pm$0.02. The correlation coefficient reaches 0.988 when dealing with the closest values. By comparing one-to-one values between our inverted \vsini\ and the median of the catalogued ones, we found an absolute mean signed difference of 5.0 \kms\ and a standard deviation of 15.0 \kms\ (Columns 4 and 5 of Tab.~\ref{accuracy}). Comparing with the closest values, the absolute mean signed difference is found to be 1.7 \kms\ with a standard deviation of 10.0 \kms\ (Columns 6 and 7 of Tab.~\ref{accuracy}). Taking into account the spread in the catalogued values by applying Eq.~\ref{deviation} decreases the average deviation to $\sim$2.0 \kms\ and the corresponding standard deviation to 9.5 \kms\ (Columns 2 and 3 of Tab.~\ref{accuracy}). This result shows that the projected equatorial rotational velocities of A-type stars can be derived with a small uncertainty using PCA-inversion techniques. 


\subsubsection{Well studied A stars}
\label{19stars}
We have also checked the deviation of our derived parameters with the ones of some well studied stars. To do that, we have selected all stars that have been studied extensively by different authors using different techniques (photometric, spectroscopic and/or interferometric). 
In the lack of a Benchmark stars list for A stars (see for instance \cite{fgkstars} for a list of FGK Benchmark stars), our selection of the well studied stars is based on the amount of references that we found and the amount of derived values for each individual parameter for each star. Stars with existing photometric determinations of \Teff/\logg\ were also taken into account in our selection. We ended up with 19 stars having more than 120 references each. The selected stars are Vega, Sirius A, HD~22484, HD~15318, HD~76644, HD~49933, HD~214994, HD~214923, HD~113139, HD~114330, HD~27819, HD~5448, HD~33256, HD~29388, 
HD~91480, HD~30210, HD~32301, HD~28355, and HD~222603.
Considering these stars, and comparing our inverted parameters to the median of the catalogued ones,  we obtain an absolute mean signed difference of 35 K with a standard deviation of 250 K for \Teff, 0.18 dex with a standard deviation of 0.40 dex for \logg, 0.09 dex with a standard deviation of 0.15 dex for [Fe/H], and 4.0 \kms\ with a standard deviation 9.5 \kms\ for \vsini. These differences and standard deviations are displayed in columns 8 and 9 of Tab.~\ref{accuracy}. The absolute mean signed differences and their corresponding standard deviations decrease when comparing with the closest catalogues values (Columns 10 and 11 of Tab.~\ref{accuracy}). These calculations also show that the effective temperatures, metallicities, and projected equatorial rotational velocities  of A stars can be derived with a good accuracy using PCA-inversion techniques.

\section{Outliers}
\label{outliers}
The outliers have been selected based on the large difference between the inverted parameters and the values retrieved from the catalogs. We have also considered as outliers, and checked in detail, stars that were found to have very low metallicities. As explained in Sec.~\ref{inversion}, gravity is usually determined with poor accuracy in spectral fitting technique for \Teff$\leq$8\,000 K, but we are confident that with our large wavelength range we are able to reach realistic values for these stars due to the sensitivity of the metallic lines to surface gravity \citep{ryabchikova2015}. This is justified by the small differences that we found between the inverted and the median \logg\ (Sec.~\ref{reference}). For that reason we have decided to exclude gravity as a parameter for the outliers selection. 

\subsection{\object{HD~6530}:}
\begin{center}
 \begin{figure}[!htp]
\includegraphics[width=\hsize]{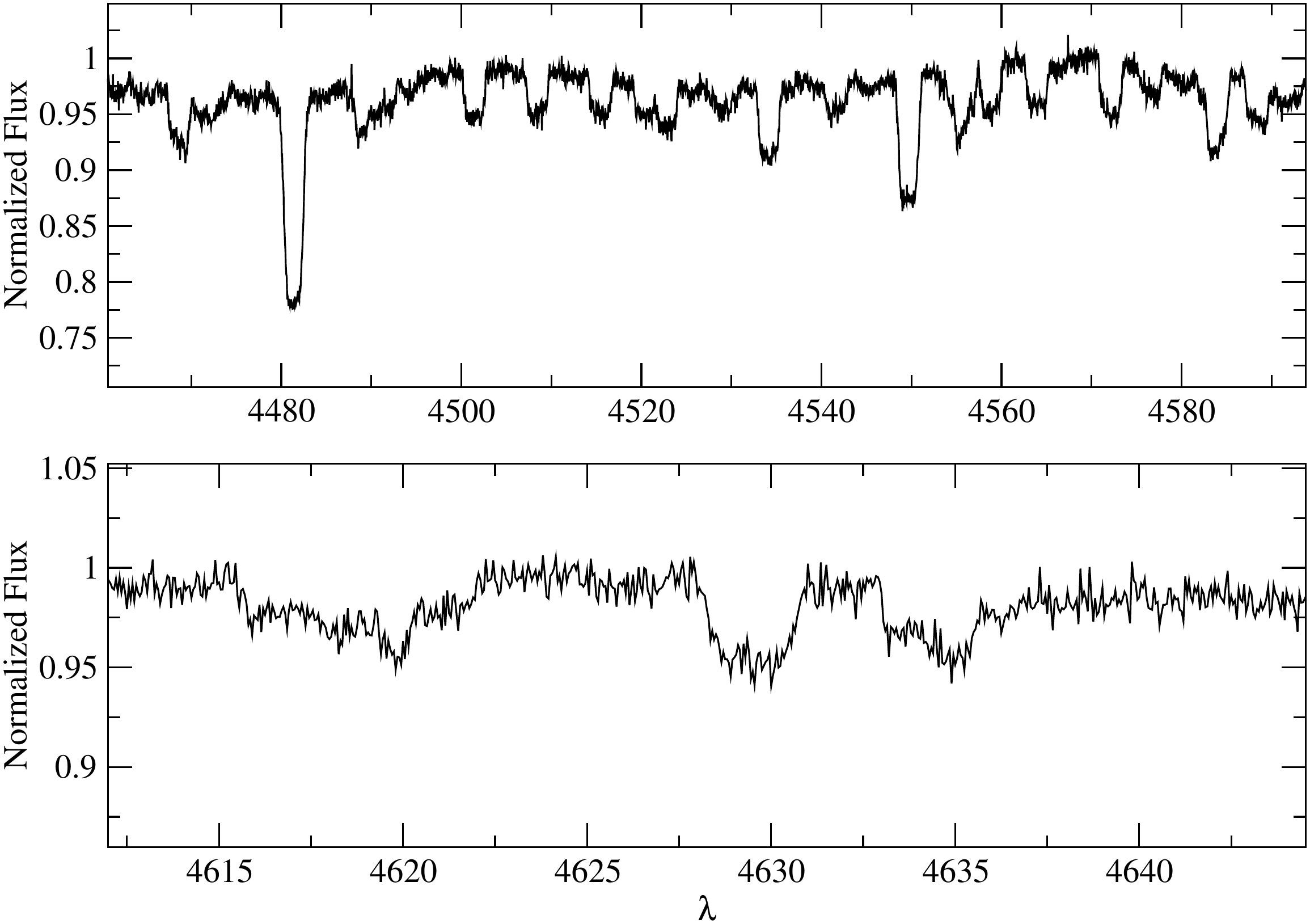}
\caption{Observed spectrum of HD~6530. Two different regions are displayed in order to show the peculiar line profile that could be due to a binary system.}
\label{HD6530}
\end{figure} 
\end{center}

\cite{cowley69} classified this star as A1V with no observable peculiarity. \cite{royer14} suspected that this star could be a binary system as the Cross-Correlation Function has an asymmetric profile. Figure~\ref{HD6530} displays the profile of several lines in the SOPHIE spectrum of HD~6530. All lines have flat cores. More observations are needed in order to confirm if HD~6530 is a binary system or whether signatures of gravity darkening due to fast rotation seen pole-on is present in the spectrum.

\subsection{\object{HD~12446}:}HD~12446 has been classified as a kA2hF2mF2 star by \cite{gray89} and is considered as a normal star in Simbad. \cite{Eggleton08} have found that this star is a member of a binary system containing a A0pSi primary (V=4.16) and an A9 secondary (V=5.27). The contribution of an Ap star and an Am star to the spectrum probably accounts for the discrepancy between the inverted parameters and those retrieved from catalogs, especially for the projected equatorial rotational velocity.

\subsection{\object{HD~16605}:}HD~16605, member of NGC~1039, is an A1-type star according to \cite{renson09}. The magnetic field of the star was discovered by \cite{Kudryavtsev} having a longitudinal component that varies from $-$2430 G to $-$840 G. \cite{balega12} considered this star as an A1p with a peculiarity of the type SiSrCr. They have also derived, spectroscopically, an effective temperature of 10\,350 K and a \vsini\ of 13 \kms\ close to our inverted values of 10\,800 K and 14 \kms, respectively. Their values were not found by querying Vizier, and hence are not present in Tab.~\ref{obs}. Using interferometric data, \cite{balega12} discovered that HD~16605 has a companion F star, 3.1 mag fainter, with an orbital period of 680 years. The fact that HD~16605 is an Ap star with overabundances of Si, Sr, and Cr could explain the large metallicity that we derived (1.1 dex). As an A1-type star, the catalog median temperature of ~8\,000 K is clearly underestimated. ATLAS9 model atmospheres and the spectrum synthesis used here do not properly account for the effect of the magnetic field. The magnetic pressure is ignored in the hydrostatic equilibrium when computing the atmospheric structure in ATLAS9 and the Zeeman splitting of the line profile is not included in the line synthesis. For that reason, the PCA-inversion technique presented in this work may not be applicable for Ap stars using ATLAS9 and SYNSPEC.

\subsection{\object{HD~23479}:} Classified as A7V, HD~23479 was suspected to be a spectroscopic binary by \cite{liu91}. It has a visual companion in the
Tycho Catalog with separation 0".85, V=9.45, and
B-V=0.54 according to \cite{malkov12}. These authors also showed that HD~23479 is a binary system containing a A7 + F6 star. Our inverted \Teff\ and \vsini\ (7300 K and 2.0 \kms) are in good agreement with previous findings (7239 K and 0.0 \kms). This star was flagged as an outlier because of the high iron abundance (+2.0 dex). This large overabundance could be explained by two contributors in the spectrum of this star, and as a spectroscopic binary, our inverted parameters do not represent the fundamental parameters of the components of HD~23479.

\subsection{\object{HD~50405}:} 
\begin{center}
 \begin{figure}[!htp]
\includegraphics[width=\hsize]{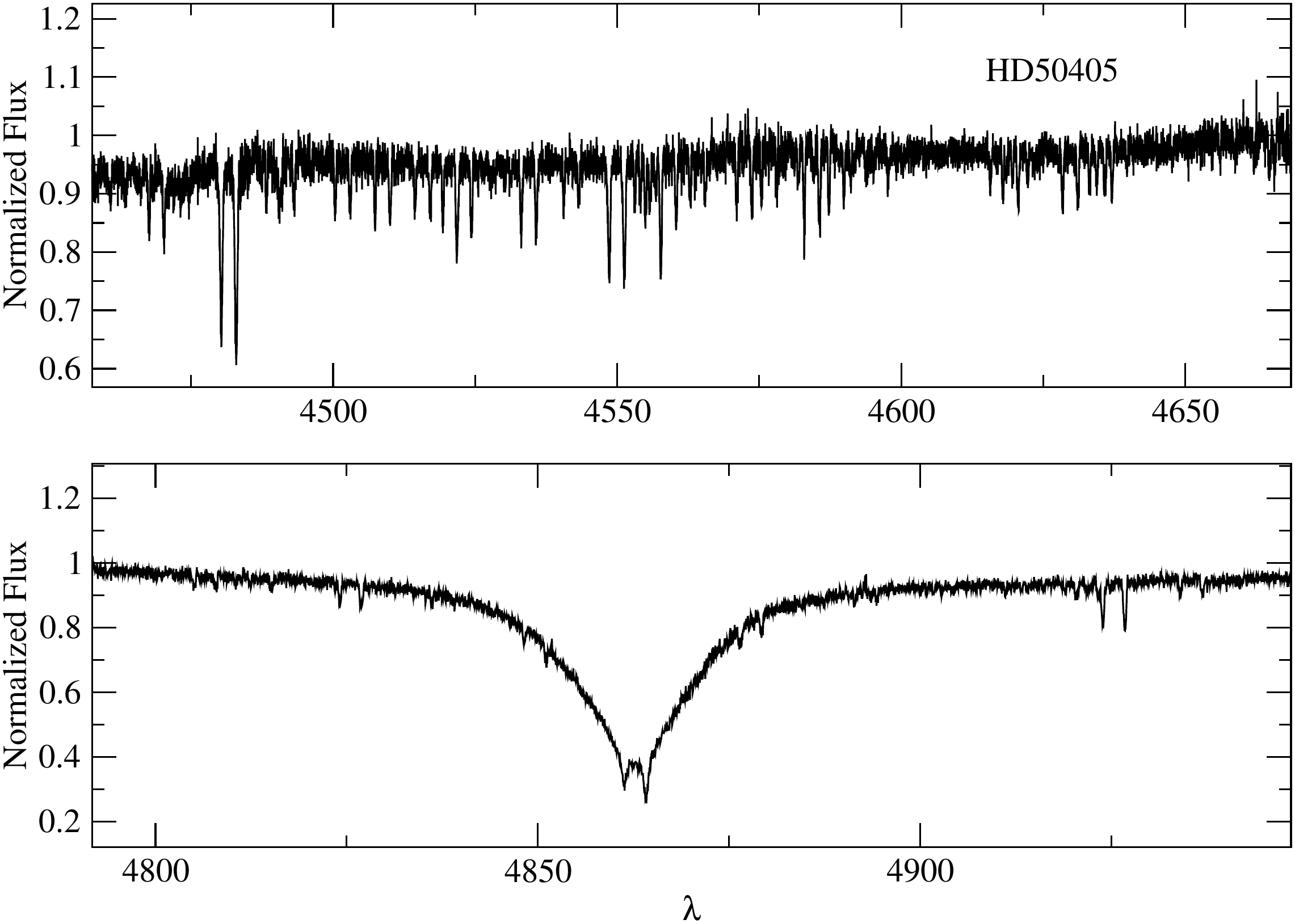}
\caption{Observed spectrum of HD~50405 around the Mg\,{\sc ii} and Fe\,{\sc ii} lines (upper panel) and the H$\beta$ line (bottom panel).}
\label{hd50405}
\end{figure} 
\end{center}

According to \cite{McCuskey56}, HD~50405 is classified as a A0 star in Simbad. \cite{lefevre09} detected, in their preliminary work, a possible variability in HD~50405 of the $\delta$ Scuti type.  By carefully inspecting the NARVAL high-resolution spectrum of HD~50405 in Fig.~\ref{hd50405}, we found a contribution of another component. This star is a member of a binary system with a component of probably a similar spectral type and should be classified as such. The variation of the radial velocity between the two components was found to be constant along the spectrum with a value close to 170 \kms. This strengthens the hypothesis that the star is actually a binary. This star will be subject to a detailed study in order to better characterize the system. The presence of double lines is the reason of the discrepancies between the inverted parameters and the catalogued ones.

\subsection{\object{HD~172271}:} According to Simbad, this star is classified as A1 by \cite{renson09}. \cite{landstreet08} detected a magnetic field of $\sim$260 G in this star and classified it as A0pCr. In their paper, they derived a \vsini\ of about 107 \kms\ very close to the 100 \kms\ that we found from our inversion. \cite{landstreet08} also derived the abundance of Cr and found an overabundance of about 2.5 dex, while Fe is about 0.5 dex overabundant. Their iron abundance is close to our inverted value of 0.4 dex. The inverted parameters of HD~172271 were rejected because it is an Ap star.

\section{Discussion and conclusion}
 \label{conclusion}
Using the Principal Component Analysis inversion, we have derived the effective temperature of "normal" A and Am stars with an average deviation of $\sim$150 K with respect to the catalogued values and with a standard deviation of 500 K.
\logg, [Fe/H] and \vsini\ are derived with a deviation of 0.35 dex, 0.15 dex and 2 \kms, respectively. Their standard deviations are 0.3 dex, 0.25 dex and 9.5 \kms, respectively. This assessment was done using an automated and objective approach that takes into account the large spread in the catalogued data.

We have also derived, for the first time, the metallicity of a large number of A-type stars. The combination of \vsini\ and [Fe/H] is a very important parameter for a first detection of the Am phenomenon. It has been shown that for the same spectral type, Am stars have lower projected equatorial rotational velocities than "normal" A-type stars and are usually slightly overabundant in iron \citep{gebran08a,gebran08b,gebran10}. 

 One could also use a learning dataset based on benchmark stars with accurate stellar parameters. This could be possible with large survey calibration such as Gaia.  The Radial Velocity Spectrometer (RVS) onboard of Gaia will collect about 15$\times 10^6$ spectra during its 5-year missions. The RVS will provide spectra in the CaII IR triplet region (from 8470 to 8710 \AA) at a spectral resolution of $\sim$11\,000. In case of A-type stars, the RVS wavelength range contains the CaII triplet as well as N~I, S~I, Si~I, and the strong Paschen hydrogen lines that are a good indicator of surface gravity at the temperature of A-type stars. \cite{recioblanco15}, using only 490 synthetic spectra as a learning database, showed that the parameters of A-type stars can be derived with high accuracy in the RVS spectral range (see Sec.~5.2 of their paper). In their work, they have compared the performance of different codes on the stellar parameters derived from the RVS spectra. We intend to test the PCA-inversion technique in a similar manner, with a much larger database, to show its efficiency at low resolution and in the RVS spectral range for not only A-type stars, but also for late and early-type stars.

A preliminary work was published in \citep{farah15} where we have estimated stellar parameters of early type stars observed in the context of the Gaia-ESO Survey (GES; \citealt{ges}). This survey collects observations of faint stars (14<V<19) taken using the GIRAFFE/FLAMES spectrograph at a medium resolution of  R$\sim$25\,000.
Two spectral regions were used, one covering the H$\delta$ line and one between 4400 and 4550 \AA, similar to the wavelength range used in the present work. Although the spectral resolution is low compared to those of the SOPHIE or PolarBase spectra, we have derived stellar parameters in good agreement with those recommended by the GES community.

As a conclusion, the work of \cite{S4n,dms} and the present one show that the PCA-inversion method proves to be fast and efficient for inverting stellar parameters of FGK and A/Am stars and deriving effective temperatures of M dwarf stars. We have built a learning database containing 2.2$\times$10$^6$ synthetic spectra at each resolution. We have used for the first time the Power Iteration algorithm to derive the eigenvectors/Principal Components of such a large database. The advantage of such an algorithm is that it can be used whatever the size of the database is. All the synthetic data have been calculated with a microturbulence velocity of 2 \kms. This parameter, in case of 1 D models, should be considered with caution, especially when attempting to model line profiles. One should modify this parameter in the learning database adopting the appropriate value for the effective temperature of the star. \cite{gebranxi} derived the dependence of $\xi_t$ on the effective temperature showing a broad maximum around 8000 K. We are working on implementing this equation in the calculations of learning databases for F-A-B stars. 

\begin{acknowledgements} 
This research has made use of the VizieR catalog access
tool, CDS, Strasbourg, France. The original description of the VizieR service
was published in A\&AS 143, 23. This research has made use of the SIMBAD
database, operated at CDS, Strasbourg, France. This research has used the Polarbase, Elodie and Sophie databases.  PolarBase is operated by the OV-GSO data center, CNRS-INSU and {\it Observatoire Midi--Pyr\'en\'ees -- Universit\'e Paul Sabatier}, Toulouse, France (Paletou et al. 2015). MG thanks Dr. Joe Malkoun for his valuable comments and help in the implementation of the Power Iteration technique. We also thank the anonymous referee for his valuable comments that helped improving this manuscript.

\end{acknowledgements}

\clearpage
\scriptsize
\onecolumn
\setcounter{table}{1}
 \begin{center}
\tablefirsthead{\hline Star & \Teff$^{inv.}$& \Teff$^{clos.}$&\Teff$^{med.}$&\logg$^{inv.}$ &\logg$^{clos.}$&\logg$^{med.}$& \vsini$^{inv.}$ & \vsini$^{clos.}$ &\vsini$^{med.}$ &\met$^{inv.}$&\met$^{clos.}$&\met$^{med.}$ & RV& Comments \\ 
& K & & & dex & & & \kms & & & dex &&& \kms& \\ \hline}
\tablehead{\hline \multicolumn{15}{|l|}{continued from previous page}\\ \hline Star & \Teff$^{inv.}$& \Teff$^{clos.}$&\Teff$^{med.}$&\logg$^{inv.}$ &\logg$^{clos.}$&\logg$^{med.}$& \vsini$^{inv.}$ & \vsini$^{clos.}$ &\vsini$^{med.}$ &\met$^{inv.}$&\met$^{clos.}$&\met$^{med.}$ & RV & Comments\\ 
& K & & & dex & & & \kms & & & dex &&& \kms& \\ \hline}
\tabletail{\hline}
\topcaption{Inverted effective temperatures, surface gravities, projected equatorial rotational velocities, metallicities, and radial velocities for the selected A stars of PolarBase. \Teff$^{clos.}$ are the values found in Vizier catalogs
closest to our inverted \Teff$^{inv.}$, and \Teff$^{med.}$ are the median of catalogs values. The same description applies to \logg, \vsini, and \met. Stars marked with the symbol (*) are the outliers (see Sec.~\ref{outliers}). SB stands for "Spectroscopic Binary", BS for "Binary star", EB for "Eclipsing Binary", Ae/Be for "Herbig Ae/Be star", and V for "Variable star". Variables include all kind of variabilities such as $\delta$ Scuti or $\gamma$ Doradus variables.}
\label{obs}
\begin{supertabular}{|l|ccc|ccc|ccc|ccc|c|c|}
 \hline
\multicolumn{15}{|c|}{\bf{PolarBase}}\\ \hline
45her & 9900.0 & 9333.0 & 8802.0 & 3.8 & 3.0 & 3.0 & 35.0 & 35.0 & 41.35 & 0.1 & 0.1 & 0.1 & -6.0&V\\
HD16605$^*$ & 10800.0 & 9601.0 & 8020.5 & 3.8 & -- & -- & 14.0 & 13.0 & 13.0 & 1.1 & -- & -- & -0.3&\\
HD25425 & 8000.0 & 8021.0 & 8500.5 & 3.1 & 4.2 & 4.2 & 14.0 & 23.3 & 26.0 & 0.3 & -- & -- & -3.5&\\
HD32633 & 10800.0 & 10500.0 & 10500.0 & 3.3 & 2.1 & 2.1 & 20.0 & 20.2 & 21.6 & 0.8 & -- & -- & -20.0&V\\
HD37357 & 7700.0 & 6346.0 & 7788.0 & 4.3 & -- & -- & 120.0 & -- & -- & -0.6 & -2.15 & -2.15 & 23.1&Ae/Be\\
HD47274 & 8900.0 & 9147.0 & 9188.5 & 3.9 & 4.1 & 4.1 & 110.0 & -- & -- & -0.2 & -- & -- & 8.0&\\
HD47485 & 7300.0 & 7157.0 & 7479.0 & 2.9 & 0.0 & 0.0 & 40.0 & -- & -- & -0.5 & 0.0 & 0.0 & 59.8&\\
HD50405$^*$ & 8100.0 & 9189.0 & 9520.0 & 3.1 & 0.0 & 0.0 & 16.0 & -- & -- & -1.8 & 0.0 & 0.0 & -55.4&\\
HD68601 & 8600.0 & 8510.0 & 7530.5 & 2.0 & -- & -- & 14.0 & 0.0 & 42.8 & 0.6 & -0.3 & -0.3 & 17.3&\\
HD84948 & 7000.0 & 6996.0 & 6996.0 & 4.4 & -- & -- & 100.0 & 55.0 & 55.0 & -0.8 & -- & -- & -59.1&SB\\
HD94194 & 7600.0 & 7614.0 & 7800.0 & 3.7 & -- & -- & 100.0 & -- & -- & -0.7 & -- & -- & -22.8&\\
HD94266 & 7100.0 & 7363.0 & 7363.0 & 2.8 & -- & -- & 90.0 & -- & -- & -0.4 & -- & -- & 1.6&\\
HD94766 & 7900.0 & 7917.0 & 8463.0 & 3.7 & -- & -- & 85.0 & -- & -- & 0.0 & -- & -- & 7.2&\\
HD96398 & 7700.0 & 7582.0 & 8606.0 & 3.0 & -- & -- & 25.0 & -- & -- & 0.6 & -- & -- & -0.4&\\
HD96399 & 7400.0 & 7400.0 & 7650.0 & 3.4 & -- & -- & 70.0 & -- & -- & -0.4 & -- & -- & 3.4&\\
HD96681 & 8200.0 & 8409.0 & 8409.0 & 3.4 & -- & -- & 80.0 & -- & -- & -0.1 & -- & -- & 19.5&\\
HD97230 & 7700.0 & 7650.0 & 7625.0 & 3.5 & 4.2 & 4.2 & 80.0 & -- & -- & 0.1 & -- & -- & 2.6&\\
HD98377 & 8800.0 & 8970.0 & 8970.0 & 3.9 & -- & -- & 50.0 & -- & -- & -0.1 & -- & -- & 1.7&\\
HD98437 & 7500.0 & 7669.0 & 7669.0 & 4.3 & -- & -- & 25.0 & -- & -- & -0.1 & -- & -- & 19.1&\\
HD98747 & 7200.0 & 7190.0 & 7190.0 & 3.2 & 4.15 & 4.15 & 35.0 & -- & -- & -0.2 & -- & -- & -12.2&\\
HD100518 & 7800.0 & 7740.0 & 7950.0 & 3.5 & -- & -- & 8.0 & 10.0 & 12.4 & -0.3 & -- & -- & 8.0&SB\\
HD100974 & 7600.0 & 7736.0 & 7736.0 & 3.5 & -- & -- & 80.0 & -- & -- & -0.4 & -- & -- & 16.0&\\
HD101470 & 7800.0 & 7716.0 & 7716.0 & 3.7 & -- & -- & 50.0 & -- & -- & -0.1 & -- & -- & -5.1&\\
HD102660 & 7300.0 & 7338.0 & 7338.0 & 3.0 & 4.15 & 4.15 & 20.0 & 21.8 & 23.9 & 0.0 & -- & -- & 0.9&SB\\
HD102841 & 7400.0 & 7400.0 & 7455.0 & 3.3 & 3.2 & 3.2 & 90.0 & -- & -- & -0.3 & -- & -- & -10.0&\\
HD103152 & 7600.0 & 7730.0 & 8059.0 & 3.4 & -- & -- & 12.0 & -- & -- & 0.0 & -- & -- & -4.9&\\
HD107612 & 8800.0 & 8970.0 & 9179.0 & 3.7 & 2.1 & 2.1 & 40.0 & 37.0 & 37.0 & 0.2 & -- & -- & 6.0&\\
HD108714 & 7800.0 & 7882.0 & 8329.5 & 3.7 & 4.2 & 4.2 & 18.0 & -- & -- & -0.8 & -- & -- & -7.8&\\
HD110932 & 11000.0 & 10833.0 & 9980.0 & 3.7 & 4.42 & 3.26 & 90.0 & 90.0 & 90.0 & -0.6 & -0.13 & -0.13 & -31.6&\\
HD112002 & 8000.0 & 7832.0 & 7832.0 & 3.5 & -- & -- & 50.0 & -- & -- & 0.1 & -- & -- & 2.2&\\
HD115403 & 7200.0 & 7150.0 & 7225.0 & 3.7 & 4.3 & 4.3 & 70.0 & -- & -- & -0.4 & -- & -- & -22.7&\\
HD116379 & 8000.0 & 8180.0 & 8190.0 & 3.8 & 4.2 & 4.2 & 80.0 & -- & -- & -0.1 & -- & -- & 1.6&\\
HD134305 & 7800.0 & 7850.0 & 8000.0 & 3.8 & 1.8 & 1.8 & 50.0 & -- & -- & 0.1 & -- & -- & -34.7&\\
HD141675 & 7700.0 & 7601.0 & 8095.0 & 3.1 & 3.77 & 3.77 & 30.0 & 40.0 & 45.0 & 0.3 & -- & -- & -2.2&\\
HD143418 & 7900.0 & 7740.0 & 8230.0 & 3.4 & 2.0 & 2.0 & 240.0 & -- & -- & 1.0 & -- & -- & -24.4&\\
HD153882 & 9200.0 & 9226.0 & 9226.0 & 3.6 & 3.5 & 3.0 & 20.0 & 23.0 & 24.5 & 0.9 & 0.61 & -0.25 & -30.1&V\\
HD155375 & 8700.0 & 8816.0 & 8657.5 & 3.8 & -- & -- & 25.0 & 25.0 & 27.85 & 0.2 & -- & -- & 21.0&V\\
HD164258 & 8600.0 & 8553.0 & 8170.0 & 3.6 & 3.58 & 3.64 & 60.0 & 60.0 & 60.0 & 0.3 & 0.31 & 0.11 & -38.5&\\
HD171948 & 10400.0 & 10500.0 & 9761.0 & 4.7 & 2.1 & 2.1 & 8.0 & 15.0 & 15.0 & -2.0 & -- & -- & 1.1&SB\\
HD172271$^*$ & 10800.0 & 9520.0 & 9145.5 & 4.1 & -- & -- & 100.0 & -- & -- & 0.4 & -- & -- & -34.5&\\
HD187547 & 7400.0 & 7500.0 & 5792.0 & 3.3 & 3.9 & 4.32 & 10.0 & 10.0 & 10.0 & -0.1 & 0.14 & 0.14 & -10.2&V\\
HD190145 & 7600.0 & 7649.0 & 7911.0 & 3.1 & -- & -- & 12.0 & -- & -- & 0.2 & -- & -- & -15.7&\\
HD192678 & 9200.0 & 9368.0 & 9511.5 & 3.4 & 4.0 & 4.0 & 10.0 & 5.0 & 5.0 & 1.5 & 1.51 & 1.51 & -39.0&V\\
HD203401 & 7600.0 & 7707.0 & 8213.5 & 3.5 & 2.0 & 2.0 & 130.0 & -- & -- & -0.8 & -- & -- & 6.0&\\
HD205117 & 9600.0 & 9609.0 & 9355.0 & 4.0 & 4.1 & 3.9 & 90.0 & 90.0 & 83.5 & -0.1 & -0.01 & 0.05 & 5.9&\\
HD245185 & 7800.0 & 5946.0 & 4823.0 & 5.0 & -- & -- & 300.0 & 80.0 & 80.0 & -2.0 & -2.98 & -2.98 & -59.1&Ae/Be\\
HD290409 & 8000.0 & 6504.0 & 6504.0 & 4.2 & -- & -- & 80.0 & -- & -- & -1.9 & -2.13 & -2.13 & 4.5&SB\\
HD322676 & 10300.0 & -- & -- & 4.5 & -- & -- & 130.0 & -- & -- & 0.1 & -- & -- & -13.3&\\
HR3320 & 7900.0 & 7910.0 & 8520.0 & 3.3 & 2.0 & 2.0 & 18.0 & 23.0 & 23.9 & 0.1 & -- & -- & 11.8&\\
Vega & 9500.0 & 9511.0 & 9520.0 & 4.0 & 4.0 & 3.98 & 25.0 & 25.0 & 23.0 & -0.5 & -0.5 & -0.53 & -14.5&\\ \hline
\multicolumn{15}{|c|}{\bf{ELODIE}}\\ \hline
alpha gem & 9900.0 & 9550.0 & 8453.0 & 4.0 & 4.0 & 4.0 & 20.0 & 25.0 & 31.5 & 0.1 & 0.03 & 0.25 & -3.3&multiple\\
gamma gem & 9000.0 & 9040.0 & 9241.0 & 3.3 & 3.53 & 3.53 & 12.0 & 13.0 & 15.0 & -0.2 & -0.15 & -0.12 & -6.0&SB\\
HD1280 & 7900.0 & 7601.0 & 8960.0 & 3.2 & 3.68 & 3.95 & 95.0 & 96.7 & 102.0 & -0.5 & 0.14 & 0.14 & 16.3&\\
HD6961 & 7500.0 & 7600.0 & 7998.0 & 3.3 & 3.55 & 3.73 & 90.0 & 97.1 & 103.0 & -0.2 & 0.0 & -1.75 & 6.8&V\\
HD7034 & 7000.0 & 7200.0 & 7294.0 & 2.3 & 4.3 & 4.3 & 95.0 & 96.0 & 96.0 & -0.4 & -- & -- & 6.5&\\
HD8538 & 7600.0 & 8020.0 & 8090.0 & 3.1 & 3.61 & 3.82 & 110.0 & 110.0 & 113.0 & -0.4 & -0.5 & -0.53 & 6.4&EB\\
HD9531 & 11000.0 & 10500.0 & 10500.0 & 3.6 & 4.0 & 4.0 & 190.0 & 195.0 & 168.5 & -0.3 & -- & -- & 23.2&V\\
HD11636 & 7800.0 & 8061.0 & 8400.0 & 3.5 & 3.73 & 3.85 & 65.0 & 65.0 & 73.0 & -0.2 & -0.1 & 0.07 & 17.0&SB\\
HD13161 & 7600.0 & 8020.0 & 8143.0 & 3.3 & 3.54 & 3.54 & 110.0 & 76.0 & 70.0 & -0.4 & -- & -- & 27.3&SB\\
HD14055 & 8000.0 & 7512.0 & 9440.0 & 3.3 & 4.1 & 4.14 & 250.0 & 254.0 & 242.0 & -1.2 & -- & -- & 6.5&\\
HD15318 & 10900.0 & 11000.0 & 10396.0 & 4.1 & 4.16 & 3.95 & 65.0 & 65.0 & 57.0 & -0.1 & -0.04 & -0.43 & 23.8&\\
HD21686 & 11000.0 & 10500.0 & 9763.5 & 4.0 & 4.1 & 4.1 & 220.0 & 244.0 & 264.5 & -0.4 & -- & -- & -13.2&\\
HD22484 & 6800.0 & 6618.0 & 5988.5 & 5.0 & 4.68 & 4.1 & 8.0 & 8.0 & 4.2 & 0.2 & 0.02 & -0.09 & 63.3&\\
HD22615 & 8100.0 & 7929.0 & 8446.0 & 3.2 & 3.3 & 4.0 & 30.0 & 30.0 & 29.75 & 0.1 & 0.34 & 0.37 & 13.3&V\\
HD23156 & 7400.0 & 7561.0 & 7705.5 & 3.4 & 4.23 & 4.5 & 30.0 & 23.65 & 42.0 & -0.3 & 0.15 & 0.15 & 32.3&V\\
HD23157 & 7100.0 & 7060.0 & 7060.0 & 3.4 & 4.26 & 4.58 & 60.0 & 77.0 & 90.0 & -0.3 & 0.12 & 0.12 & 26.5&\\
HD23194 & 7800.0 & 8031.0 & 8253.5 & 3.7 & 4.0 & 4.0 & 35.0 & 30.4 & 20.0 & -0.4 & -0.2 & -0.17 & 27.2&V\\
HD23325 & 7200.0 & 7000.0 & 7531.0 & 3.1 & 4.23 & 4.23 & 75.0 & 75.0 & 75.0 & -0.3 & 0.34 & 0.34 & 27.2&\\
HD23351 & 6800.0 & 6755.0 & 6622.0 & 4.6 & 4.38 & 4.38 & 16.0 & 70.0 & 75.0 & -0.2 & 0.06 & 0.26 & 45.4&\\
HD23375 & 7000.0 & 7336.0 & 7387.5 & 3.4 & 4.22 & 4.22 & 85.0 & 75.0 & 75.0 & -0.4 & 0.15 & 0.15 & 23.0&V\\
HD23607 & 7300.0 & 7586.0 & 7780.0 & 3.2 & 3.97 & 4.0 & 18.0 & 10.0 & 10.0 & -0.3 & -0.3 & -0.17 & 33.1&V\\
HD23763 & 7800.0 & 8079.0 & 8431.5 & 3.5 & 4.1 & 4.1 & 110.0 & 107.0 & 100.0 & -0.8 & -0.34 & -0.34 & 19.8&\\
HD23850 & 11000.0 & 10540.0 & 12200.0 & 2.9 & -- & -- & 170.0 & 170.0 & 195.0 & -1.2 & -- & -- & 23.6&SB\\
HD23924 & 7600.0 & 7750.0 & 8000.0 & 3.5 & 3.94 & 4.0 & 30.0 & 44.8 & 80.0 & -0.2 & -0.3 & -0.3 & 33.0&\\
HD27749 & 7300.0 & 7289.0 & 7369.0 & 3.4 & 3.0 & 4.15 & 14.0 & 14.0 & 14.2 & 0.3 & 0.4 & 0.52 & 29.2&SB\\
HD27962 & 8300.0 & 8603.0 & 8850.0 & 3.4 & 3.8 & 3.8 & 12.0 & 12.7 & 13.0 & -0.1 & 0.13 & 0.13 & 63.6&V\\
HD28319 & 7600.0 & 7650.0 & 7830.0 & 4.0 & 3.7 & 3.59 & 85.0 & 87.5 & 78.0 & -0.1 & 0.01 & 0.01 & 46.6&V\\
HD28355 & 7600.0 & 7600.0 & 7850.0 & 4.0 & 4.0 & 4.1 & 90.0 & 92.8 & 105.0 & 0.2 & 0.35 & 0.35 & 27.1&\\
HD28527 & 7700.0 & 7750.0 & 8180.0 & 3.7 & 3.98 & 4.1 & 70.0 & 70.0 & 86.0 & 0.1 & 0.13 & 0.13 & 27.2&V\\
HD28546 & 7400.0 & 7490.0 & 7626.0 & 3.6 & 3.85 & 4.11 & 25.0 & 28.2 & 30.45 & 0.1 & 0.13 & 0.13 & 72.7&SB\\
HD28978 & 8400.0 & 8306.0 & 8996.5 & 3.1 & 3.42 & 3.7 & 25.0 & 24.0 & 21.4 & -0.3 & -0.26 & 0.14 & 23.4&\\
HD30453 & 7300.0 & 7362.0 & 7811.0 & 2.8 & -- & -- & 12.0 & 15.0 & 23.0 & -0.1 & -- & -- & 94.8&SB\\
HD30739 & 7500.0 & 7199.0 & 9392.0 & 2.6 & 3.9 & 4.0 & 240.0 & 235.0 & 212.0 & -1.2 & -0.2 & -0.2 & 3.3&\\
HD31295 & 7900.0 & 7897.0 & 8968.0 & 3.8 & 4.0 & 4.09 & 130.0 & 121.0 & 116.5 & -1.5 & -1.31 & -0.85 & -12.0&\\
HD32537 & 6800.0 & 6699.0 & 6978.0 & 3.6 & 3.9 & 4.03 & 18.0 & 19.0 & 19.8 & -0.4 & -0.36 & -0.3 & 16.3&V\\
HD33111 & 7500.0 & 7330.0 & 8043.0 & 3.0 & 3.4 & 3.7 & 180.0 & 180.0 & 190.0 & -0.6 & -0.2 & -0.2 & 3.2&\\
HD33256 & 6800.0 & 7000.0 & 6406.0 & 5.0 & 4.5 & 4.05 & 10.0 & 10.0 & 9.7 & -0.2 & -0.2 & -0.3 & 40.2&\\
HD33641 & 7500.0 & 7560.0 & 8200.0 & 3.4 & 3.92 & 3.92 & 80.0 & 84.0 & 85.5 & -0.3 & -- & -- & 23.8&\\
HD33959 & 7200.0 & 6966.0 & 7670.0 & 2.7 & 3.46 & 3.53 & 25.0 & 24.5 & 27.8 & -0.4 & -0.14 & -0.01 & -9.9&V\\
HD38090 & 7400.0 & 7321.0 & 8491.0 & 2.0 & 4.2 & 4.2 & 140.0 & 180.0 & 204.0 & -0.9 & -- & -- & 44.3&\\
HD40536 & 7600.0 & 7740.0 & 8137.5 & 3.3 & -- & -- & 12.0 & 48.0 & 50.55 & -0.1 & 0.38 & 0.38 & 99.9&SB\\
HD40932 & 7600.0 & 7447.0 & 8079.5 & 3.1 & 3.7 & 4.0 & 12.0 & 14.2 & 18.0 & -0.3 & -0.1 & -0.1 & 34.0&SB\\
HD43378 & 8400.0 & 8454.0 & 9120.0 & 3.5 & 4.05 & 4.17 & 45.0 & 45.0 & 45.0 & -0.6 & -0.32 & -0.32 & 6.5&EB\\
HD49933 & 6800.0 & 6751.0 & 6580.0 & 4.9 & 4.5 & 4.2 & 10.0 & 9.9 & 10.45 & -0.5 & -0.49 & -0.44 & 9.9&\\
HD58142 & 10000.0 & 9692.0 & 9500.0 & 3.8 & 3.82 & 3.96 & 18.0 & 18.0 & 18.6 & 0.0 & -- & -- & 45.9&\\
HD61273 & 7200.0 & 5904.0 & 5720.0 & 5.0 & -- & -- & 40.0 & -- & -- & -0.4 & -- & -- & 76.8&V\\
HD61421 & 6800.0 & 6704.0 & 6605.0 & 4.5 & 4.43 & 4.02 & 8.0 & 10.0 & 5.54 & 0.0 & 0.0 & -0.02 & -3.0&SB\\
HD62454 & 6800.0 & 6920.0 & 7052.0 & 3.5 & 4.04 & 4.1 & 14.0 & -- & -- & -0.5 & 0.1 & 0.13 & 56.0&V\\
HD72660 & 9200.0 & 9226.0 & 9606.0 & 3.6 & 3.39 & 4.0 & 6.0 & 5.0 & 9.0 & 0.1 & 0.21 & 0.28 & -3.3&\\
HD72942 & 7800.0 & 8130.0 & 8720.0 & 3.3 & 3.8 & 3.8 & 75.0 & 67.5 & 61.9 & -0.3 & -0.05 & 0.05 & 23.8&SB\\
HD73045 & 7100.0 & 7199.0 & 7524.0 & 2.9 & 4.0 & 4.0 & 14.0 & 12.1 & 10.8 & 0.1 & 0.35 & 0.41 & 24.5&SB\\
HD73174 & 7800.0 & 8090.0 & 8200.0 & 3.2 & 4.0 & 4.0 & 8.0 & 8.1 & 11.1 & 0.2 & 0.3 & 0.3 & 9.7&SB\\
HD73666 & 9400.0 & 9503.0 & 8356.5 & 3.7 & -- & -- & 12.0 & 17.1 & 20.0 & 0.1 & 0.6 & 0.62 & 27.3&BS\\
HD73709 & 7700.0 & 8060.0 & 4966.0 & 2.9 & 2.67 & 2.63 & 16.0 & 16.8 & 29.8 & 0.2 & 0.17 & 0.23 & 0.0&SB\\
HD73730 & 7600.0 & 8020.0 & 8200.0 & 3.1 & 4.0 & 4.0 & 30.0 & 30.0 & 30.2 & 0.0 & 0.4 & 0.4 & 27.2&V\\
HD73731 & 7600.0 & 7851.0 & 8100.0 & 3.2 & -- & -- & 65.0 & 63.3 & 62.15 & -0.1 & -- & -- & 6.6&SB\\
HD77350 & 10700.0 & 10551.0 & 10325.0 & 3.8 & 3.6 & 3.59 & 20.0 & 20.0 & 18.0 & 0.2 & 0.15 & 0.1 & -27.3&SB\\
HD88983 & 7600.0 & 7450.0 & 7923.5 & 3.4 & 4.01 & 4.01 & 110.0 & 105.0 & 118.0 & -0.5 & -0.2 & -0.2 & 6.7&\\
HD89774 & 10700.0 & 9500.0 & 9230.0 & 4.1 & 4.1 & 4.1 & 70.0 & 70.0 & 60.0 & 0.2 & -- & -- & 26.5&V\\
HD89822 & 11000.0 & 10500.0 & 9677.0 & 4.1 & 3.95 & 3.85 & 8.0 & 10.0 & 15.0 & 0.1 & 0.13 & 0.15 & 33.6&V\\
HD91480 & 6800.0 & 6868.0 & 7000.0 & 3.7 & 4.21 & 4.3 & 40.0 & 36.9 & 52.2 & -0.3 & -0.19 & -0.11 & -3.3&\\
HD95418 & 8300.0 & 8734.0 & 9470.0 & 3.1 & 3.82 & 3.96 & 40.0 & 40.0 & 41.0 & -0.7 & -0.76 & 0.01 & -10.0&V\\
HD95608 & 8300.0 & 7577.0 & 9250.0 & 3.6 & 4.1 & 4.16 & 16.0 & 15.4 & 21.0 & -0.2 & -- & -- & -13.7&V\\
HD97633 & 8800.0 & 8790.0 & 9302.0 & 3.3 & 3.45 & 3.6 & 20.0 & 20.0 & 22.0 & -0.2 & -0.16 & 0.04 & -13.2&V\\
HD98664 & 11000.0 & 10680.0 & 10618.0 & 3.9 & 3.89 & 3.99 & 65.0 & 65.0 & 57.0 & -0.2 & -- & -- & 10.1&\\
HD99285 & 6800.0 & 6810.0 & 6754.0 & 3.9 & 3.92 & 3.88 & 35.0 & 34.4 & 35.2 & -0.3 & -0.34 & -0.19 & 12.2&\\
HD99859 & 7600.0 & 7986.0 & 8596.0 & 3.5 & 4.29 & 3.14 & 90.0 & 90.0 & 84.0 & -0.4 & -- & -- & 23.4&\\
HD106103 & 6800.0 & 6737.0 & 6708.5 & 4.9 & 4.5 & 4.31 & 20.0 & 18.8 & 10.0 & -0.2 & -0.29 & -0.02 & 19.9&V\\
HD106251 & 7200.0 & 7415.0 & 7914.0 & 3.0 & 2.0 & 3.11 & 35.0 & 45.0 & 50.4 & -0.1 & -- & -- & 27.3&\\
HD106293 & 6800.0 & 6685.0 & 6568.0 & 5.0 & 4.34 & 4.34 & 45.0 & -- & -- & -0.2 & -0.2 & 0.06 & 18.9&\\
HD106691 & 6800.0 & 6739.0 & 6718.5 & 5.0 & 4.5 & 4.38 & 35.0 & 30.0 & 28.7 & -0.2 & -0.24 & -0.06 & 19.9&\\
HD106887 & 7900.0 & 8291.0 & 8400.0 & 3.8 & 4.2 & 4.2 & 75.0 & 75.0 & 84.1 & 0.0 & 0.21 & 0.21 & 6.8&\\
HD106946 & 6800.0 & 6820.0 & 6894.5 & 4.4 & 4.46 & 4.39 & 60.0 & 52.9 & 51.45 & -0.2 & -0.18 & -0.06 & 16.8&\\
HD106999 & 7800.0 & 8116.0 & 8148.0 & 3.7 & 4.09 & 4.09 & 40.0 & 51.4 & 55.7 & -0.1 & 0.08 & 0.22 & 12.2&\\
HD107131 & 7700.0 & 7750.0 & 7895.0 & 4.6 & 4.21 & 4.2 & 190.0 & 190.0 & 187.0 & -0.1 & -- & -- & 23.1&V\\
HD107168 & 8000.0 & 8026.0 & 8200.0 & 3.3 & 4.15 & 4.18 & 14.0 & 16.3 & 16.3 & 0.4 & 0.39 & 0.38 & 18.2&\\
HD107276 & 7700.0 & 7710.0 & 7969.0 & 4.4 & 4.21 & 4.21 & 110.0 & 95.0 & 95.0 & 0.0 & -0.05 & -0.05 & 16.8&\\
HD107513 & 7100.0 & 7212.0 & 7360.0 & 3.8 & 3.99 & 4.25 & 65.0 & 55.6 & 50.0 & -0.1 & -0.05 & -0.05 & 21.2&V\\
HD107655 & 8200.0 & 7638.0 & 9520.0 & 3.4 & 4.1 & 4.1 & 50.0 & 50.0 & 46.0 & -0.7 & 0.08 & 0.08 & 10.2&\\
HD107966 & 8000.0 & 7999.0 & 8600.0 & 3.3 & 3.73 & 3.82 & 50.0 & 51.0 & 51.0 & -0.4 & -0.13 & 0.05 & 27.0&V\\
HD108382 & 7900.0 & 8282.0 & 8370.0 & 3.4 & 3.69 & 3.92 & 75.0 & 74.0 & 80.0 & -0.3 & -0.14 & -0.05 & 10.0&\\
HD108486 & 7600.0 & 7678.0 & 7844.5 & 3.4 & 4.19 & 4.35 & 35.0 & 33.2 & 30.0 & -0.3 & 0.23 & 0.23 & 16.7&EB\\
HD108642 & 7500.0 & 7604.0 & 7834.0 & 3.1 & -- & -- & 8.0 & 9.6 & 11.5 & -0.2 & -0.21 & -0.08 & 44.2&SB\\
HD108651 & 7500.0 & 7694.0 & 7909.0 & 3.3 & 4.24 & 4.5 & 20.0 & 18.0 & 18.0 & 0.0 & 0.4 & 0.44 & 29.7&SB\\
HD108662 & 10000.0 & 9980.0 & 9903.5 & 3.7 & 3.4 & 3.4 & 25.0 & 20.0 & 16.1 & 1.0 & 0.02 & 0.02 & -20.0&V\\
HD108945 & 8300.0 & 8327.0 & 8853.0 & 3.3 & 3.2 & 3.68 & 65.0 & 63.8 & 62.9 & 0.0 & -0.1 & 0.01 & 9.0&V\\
HD109307 & 8000.0 & 8162.0 & 8414.0 & 3.3 & 3.91 & 4.13 & 14.0 & 14.8 & 18.0 & -0.2 & -0.05 & -0.05 & 16.6&\\
HD109530 & 6800.0 & 6890.0 & 6470.0 & 5.0 & 3.71 & 3.71 & 65.0 & -- & -- & -0.1 & -0.04 & 0.06 & -3.4&\\
HD113139 & 6800.0 & 6794.0 & 6859.5 & 4.1 & 4.13 & 4.26 & 85.0 & 85.6 & 91.85 & -0.2 & -0.05 & 0.08 & 10.0&\\
HD114330 & 9400.0 & 9509.0 & 9509.0 & 3.3 & 3.6 & 3.8 & 6.0 & 8.6 & 15.0 & -0.1 & -0.13 & -0.02 & -23.3&\\
HD116706 & 8100.0 & 8182.0 & 8720.0 & 3.5 & 4.14 & 4.17 & 55.0 & 54.0 & 54.0 & -0.4 & -- & -- & 22.1&\\
HD118022 & 9200.0 & 9238.0 & 9465.0 & 3.6 & 3.5 & 3.5 & 14.0 & 15.0 & 19.0 & 1.1 & 1.14 & 0.73 & -31.0&V\\
HD119537 & 7900.0 & 7544.0 & 9066.0 & 3.0 & 4.1 & 4.1 & 14.0 & 13.0 & 13.0 & -0.9 & -- & -- & 82.2&\\
HD122408 & 7500.0 & 7427.0 & 8299.0 & 2.9 & 3.58 & 3.6 & 160.0 & 165.0 & 166.5 & -0.8 & -0.27 & -0.27 & 19.2&\\
HD126661 & 7400.0 & 7400.0 & 7570.5 & 2.6 & 1.7 & 3.59 & 30.0 & 30.0 & 32.4 & 0.0 & 0.1 & 0.36 & -10.0&\\
HD127304 & 10600.0 & 10500.0 & 9790.0 & 4.2 & 4.1 & 4.1 & 8.0 & 7.4 & 14.0 & 0.2 & -- & -- & -13.0&\\
HD142908 & 6800.0 & 6848.0 & 6930.0 & 3.6 & 3.95 & 3.98 & 75.0 & 75.7 & 75.7 & -0.2 & -0.14 & -0.02 & 3.2&\\
HD143807 & 11000.0 & 11224.0 & 10080.0 & 3.9 & 3.84 & 3.72 & 4.0 & 10.0 & 12.0 & 0.0 & 0.0 & -0.0 & -13.4&SB\\
HD145389 & 11000.0 & 11455.0 & 11455.0 & 3.7 & 3.81 & 3.82 & 14.0 & 15.0 & 15.0 & 0.0 & 0.08 & 0.23 & -6.8&V\\
HD145647 & 8300.0 & 8799.0 & 9488.5 & 3.2 & 4.1 & 4.1 & 45.0 & 43.0 & 43.0 & -0.8 & -- & -- & -3.3&\\
HD152107 & 8300.0 & 8200.0 & 8419.5 & 3.3 & 4.04 & 4.07 & 25.0 & 28.5 & 33.0 & 0.3 & 1.65 & 1.65 & 3.4&V\\
HD153808 & 7700.0 & 7626.0 & 9390.0 & 2.9 & 3.92 & 3.92 & 210.0 & 85.0 & 60.0 & -0.8 & -0.25 & -0.25 & -90.0&SB\\
HD158716 & 8600.0 & 8629.0 & 9066.0 & 3.6 & 4.1 & 4.17 & 6.0 & 15.0 & 18.5 & -0.1 & -0.04 & -0.04 & -23.2&\\
HD176984 & 10400.0 & 11000.0 & 9193.0 & 3.6 & -- & -- & 30.0 & 25.3 & 23.0 & 0.0 & -- & -- & -56.8&\\
HD200761 & 10700.0 & 10782.0 & 9616.0 & 4.3 & 4.1 & 4.01 & 110.0 & 104.0 & 104.0 & 0.2 & 0.26 & 0.26 & 12.4&\\
HD209459 & 11000.0 & 11000.0 & 10465.5 & 3.7 & 3.55 & 3.5 & 4.0 & 4.0 & 14.0 & -0.3 & -0.15 & -0.11 & 29.6&\\
HD214923 & 11000.0 & 10759.0 & 11430.0 & 3.5 & 3.75 & 3.97 & 180.0 & 185.0 & 163.0 & -0.3 & -0.1 & 0.03 & 20.3&\\
HD218045 & 11000.0 & 10616.0 & 10220.0 & 4.0 & 3.98 & 3.7 & 130.0 & 131.0 & 144.0 & 0.0 & 0.0 & -0.23 & 9.8&V\\
Sirius A & 9800.0 & 9870.0 & 9870.0 & 4.2 & 4.27 & 4.30 & 18.0 & 16.0 & 15.0 & 0.2 & 0.28 & 0.36 & 0.0&\\\hline
\hline
\multicolumn{15}{|c|}{\bf{SOPHIE}}\\ \hline
HD1439 & 10100.0 & 9790.0 & 9491.5 & 3.8 & 4.1 & 4.1 & 50.0 & 50.0 & 39.0 & 0.1 & -- & -- & -12.2&\\
HD1561 & 10100.0 & 9790.0 & 9375.0 & 3.8 & 4.1 & 4.1 & 70.0 & 60.0 & 60.0 & 0.1 & -- & -- & -19.9&\\
HD4058 & 7800.0 & 7936.0 & 8296.0 & 4.1 & 3.96 & 3.96 & 210.0 & 60.0 & 58.0 & 0.0 & -- & -- & -90.0&SB\\
HD5448 & 8000.0 & 7981.0 & 8050.5 & 3.3 & 3.78 & 3.86 & 70.0 & 69.3 & 75.0 & -0.1 & -0.19 & -0.04 & 3.3&\\
HD6530$^*$ & 10400.0 & 9520.0 & 9485.0 & 4.2 & 4.1 & 4.1 & 100.0 & 70.0 & 51.0 & 0.3 & -- & -- & -6.1&\\
HD6695 & 8200.0 & 8266.0 & 8587.5 & 3.7 & 3.69 & 4.05 & 150.0 & 149.0 & 143.0 & -0.2 & -0.46 & -0.91 & -13.6&\\
HD6813 & 7900.0 & 8047.0 & 8721.0 & 3.4 & 2.1 & 2.1 & 70.0 & -- & -- & -0.3 & -- & -- & -19.8&\\
HD6961 & 7800.0 & 7787.0 & 7998.0 & 3.3 & 3.55 & 3.73 & 90.0 & 97.1 & 103.0 & 0.0 & 0.0 & -1.75 & 3.1&V\\
HD8374 & 7200.0 & 7275.0 & 7370.0 & 3.1 & 3.9 & 3.9 & 30.0 & 30.0 & 32.9 & 0.1 & -- & -- & 13.1&SB\\
HD12111 & 7700.0 & 7872.0 & 8538.0 & 3.7 & 4.17 & 4.17 & 60.0 & 67.0 & 71.6 & -0.4 & -- & -- & -12.4&SB\\
HD12446$^*$ & 7900.0 & 8200.0 & 8650.0 & 2.5 & 3.99 & 3.99 & 140.0 & 84.0 & 70.8 & -0.2 & -- & -- & -6.7&\\
HD13041 & 7900.0 & 8180.0 & 8289.0 & 3.3 & 3.65 & 3.88 & 130.0 & 133.0 & 133.0 & -0.4 & -0.5 & -0.98 & 3.4&\\
HD14252 & 8800.0 & 8970.0 & 9000.0 & 3.4 & 3.75 & 3.97 & 25.0 & 25.0 & 22.0 & -0.1 & 0.33 & 0.33 & -3.0&\\
HD15082 & 7100.0 & 7268.0 & 7354.0 & 2.9 & 2.0 & 4.3 & 80.0 & 90.0 & 90.0 & -0.2 & 0.1 & 0.44 & -3.3&V\\
HD17471 & 10900.0 & 10500.0 & 9708.5 & 4.0 & 4.1 & 4.1 & 60.0 & 54.0 & 54.0 & 0.1 & -- & -- & 6.6&V\\
HD20149 & 9800.0 & 9500.0 & 9228.0 & 3.8 & 4.1 & 4.1 & 25.0 & 23.0 & 23.0 & 0.0 & -- & -- & -13.2&\\
HD20677 & 8300.0 & 8475.0 & 8785.0 & 3.8 & 4.2 & 4.22 & 130.0 & 134.0 & 144.0 & -0.6 & -- & -- & -22.8&V\\
HD21050 & 10700.0 & 10142.0 & 9712.5 & 4.2 & 4.25 & 4.18 & 30.0 & 27.0 & 27.0 & -0.1 & -- & -- & -3.4&\\
HD23190 & 7400.0 & 7300.0 & 7848.0 & 3.3 & 4.2 & 4.3 & 30.0 & -- & -- & -0.3 & 0.21 & 0.21 & 3.4&\\
HD23247 & 6800.0 & 6811.0 & 6432.5 & 3.9 & -- & -- & 40.0 & 40.0 & 40.0 & -0.3 & -0.41 & 0.04 & 6.7&\\
HD23387 & 7900.0 & 7865.0 & 7807.5 & 3.3 & 4.0 & 4.0 & 25.0 & 23.5 & 21.3 & -1.3 & -0.38 & -0.34 & 0.0&\\
HD23441 & 10900.0 & 10500.0 & 9938.0 & 4.0 & 4.1 & 4.1 & 240.0 & 250.0 & 232.0 & -0.1 & -- & -- & -17.0&\\
HD23479$^*$ & 7300.0 & 7239.0 & 6994.5 & 4.9 & -- & -- & 2.0 & 0.0 & 75.0 & 2.0 & -- & -- & -6.1&\\
HD23489 & 8500.0 & 8500.0 & 9000.0 & 3.8 & 4.25 & 4.25 & 120.0 & 110.0 & 105.0 & -0.4 & -0.02 & -0.02 & -10.1&\\
HD23791 & 7300.0 & 7250.0 & 7776.0 & 3.3 & 4.32 & 4.32 & 70.0 & 72.8 & 85.0 & -0.1 & 0.15 & 0.15 & 3.3&\\
HD23863 & 7600.0 & 7750.0 & 7880.0 & 3.5 & 4.1 & 4.1 & 140.0 & 137.0 & 160.0 & -0.4 & 0.0 & 0.01 & -6.8&\\
HD24368 & 8900.0 & 8970.0 & 8603.5 & 3.4 & 4.0 & 4.1 & 10.0 & 25.7 & 30.0 & 0.0 & 0.2 & 0.2 & 6.5&V\\
HD25175 & 10600.0 & 10500.0 & 8848.0 & 3.9 & -- & -- & 60.0 & 55.0 & 55.0 & 0.0 & -- & -- & 26.7&\\
HD25490 & 8300.0 & 8478.0 & 9017.0 & 3.5 & 4.06 & 4.08 & 80.0 & 83.0 & 76.1 & -0.6 & -0.22 & -0.22 & -12.8&\\
HD25642 & 10900.0 & 10585.0 & 9482.0 & 3.7 & 4.1 & 4.1 & 210.0 & 205.0 & 200.5 & 0.2 & 0.01 & 0.01 & -13.6&\\
HD26764 & 7900.0 & 8214.0 & 9000.0 & 3.1 & 4.2 & 4.2 & 220.0 & 228.0 & 230.0 & -0.7 & -- & -- & -29.3&\\
HD27628 & 6900.0 & 7000.0 & 7304.0 & 2.8 & 3.93 & 4.0 & 30.0 & 30.0 & 30.0 & 0.0 & 0.14 & 0.17 & 44.1&V\\
HD27819 & 7800.0 & 7850.0 & 7997.0 & 3.5 & 3.89 & 4.15 & 40.0 & 41.0 & 42.0 & -0.1 & -0.06 & 0.14 & 37.1&\\
HD27934 & 7900.0 & 7850.0 & 8250.0 & 3.4 & 3.8 & 3.83 & 80.0 & 81.0 & 86.0 & -0.1 & 0.05 & 0.05 & 30.6&V\\
HD27946 & 7300.0 & 7360.0 & 7592.0 & 3.6 & 4.2 & 4.2 & 150.0 & 145.0 & 191.0 & -0.3 & -- & -- & 12.3&V\\
HD27962 & 8900.0 & 8850.0 & 8850.0 & 3.7 & 3.8 & 3.8 & 12.0 & 12.7 & 13.0 & 0.3 & 0.2 & 0.13 & 36.2&V\\
HD28024 & 7000.0 & 7150.0 & 7449.0 & 2.7 & 3.72 & 4.01 & 200.0 & 196.0 & 215.0 & -0.4 & -- & -- & 3.1&V\\
HD28226 & 7200.0 & 7215.0 & 7600.5 & 3.1 & 4.09 & 4.09 & 80.0 & 90.0 & 99.0 & 0.0 & 0.31 & 0.31 & 33.4&\\
HD28527 & 7800.0 & 7757.0 & 8180.0 & 3.4 & 3.98 & 4.1 & 60.0 & 70.0 & 86.0 & 0.0 & 0.13 & 0.13 & 33.5&V\\
HD28910 & 7300.0 & 7276.0 & 7640.0 & 3.2 & 3.81 & 3.91 & 100.0 & 95.0 & 126.0 & -0.2 & 0.12 & 0.12 & 33.2&V\\
HD29388 & 8000.0 & 7928.0 & 8380.0 & 3.4 & 3.88 & 4.04 & 80.0 & 80.0 & 89.0 & -0.1 & 0.0 & 0.06 & 39.8&\\
HD29488 & 7800.0 & 7800.0 & 8146.0 & 3.2 & 3.97 & 4.09 & 120.0 & 120.0 & 128.0 & -0.2 & 0.1 & 0.1 & 23.0&\\
HD29499 & 7300.0 & 7400.0 & 7810.0 & 3.0 & 2.0 & 4.02 & 60.0 & 60.0 & 70.0 & 0.1 & 0.36 & 0.36 & 39.4&\\
HD30210 & 7700.0 & 7694.0 & 8017.0 & 3.2 & 3.66 & 3.8 & 50.0 & 45.0 & 57.2 & 0.2 & 0.13 & 0.13 & 33.9&\\
HD30780 & 7300.0 & 7600.0 & 7763.0 & 3.1 & 3.88 & 3.9 & 150.0 & 145.0 & 145.0 & -0.3 & 0.21 & 0.21 & 13.6&V\\
HD32297 & 7500.0 & 7942.0 & 7978.5 & 3.8 & 2.0 & 2.0 & 90.0 & -- & -- & -0.7 & -0.44 & -0.44 & 10.0&\\
HD32301 & 7800.0 & 7850.0 & 8113.0 & 3.2 & 3.73 & 3.88 & 110.0 & 115.0 & 125.5 & -0.1 & 0.15 & 0.15 & 27.2&\\
HD33204 & 7300.0 & 7530.0 & 7646.0 & 2.9 & 4.0 & 4.06 & 35.0 & 34.5 & 35.3 & 0.1 & 0.35 & 0.35 & 42.0&EB\\
HD33254 & 7500.0 & 7522.0 & 7860.0 & 3.2 & 4.04 & 4.12 & 14.0 & 14.1 & 18.5 & 0.3 & 0.45 & 0.45 & 45.1&SB\\
HD42818 & 10900.0 & 10834.0 & 9520.0 & 4.4 & 4.16 & 4.13 & 260.0 & 255.0 & 255.0 & 0.3 & -- & -- & -30.7&\\
HD43378 & 9500.0 & 9580.0 & 9120.0 & 4.0 & 4.05 & 4.17 & 50.0 & 47.0 & 45.0 & 0.1 & -0.27 & -0.32 & -9.3&EB\\
HD44769 & 7500.0 & 7400.0 & 7717.0 & 3.3 & 3.65 & 3.81 & 110.0 & 124.0 & 135.0 & -0.3 & -0.24 & -0.24 & 6.1&\\
HD47105 & 9100.0 & 9049.0 & 9241.0 & 3.4 & 3.53 & 3.53 & 12.0 & 13.0 & 15.0 & -0.1 & -0.12 & -0.12 & -13.2&SB\\
HD48097 & 8500.0 & 8509.0 & 8985.0 & 3.8 & 4.2 & 4.27 & 110.0 & 102.0 & 100.0 & -0.3 & -- & -- & 3.4&\\
HD49908 & 10200.0 & 10386.0 & 9110.5 & 3.9 & 4.2 & 4.2 & 110.0 & 117.0 & 117.0 & 0.2 & -- & -- & 16.7&\\
HD50747 & 7500.0 & 7527.0 & 8051.0 & 3.2 & 4.2 & 2.1 & 50.0 & 61.0 & 67.0 & -0.4 & -0.32 & -0.32 & -9.7&\\
HD51106 & 7500.0 & 7638.0 & 8027.0 & 3.8 & 3.88 & 2.99 & 20.0 & -- & -- & -0.2 & 0.37 & 0.37 & 0.0&\\
HD70313 & 8000.0 & 8077.0 & 8660.0 & 3.7 & 4.2 & 4.24 & 110.0 & 112.0 & 112.0 & -0.3 & -- & -- & 3.4&\\
HD72846 & 7800.0 & 7800.0 & 7995.5 & 3.4 & 4.2 & 4.2 & 110.0 & 107.0 & 116.0 & -0.2 & -- & -- & 18.1&\\
HD73210 & 7400.0 & 7400.0 & 7866.0 & 3.0 & -- & -- & 90.0 & 81.4 & 75.0 & -0.4 & -- & -- & 30.0&SB\\
HD73262 & 11000.0 & 11055.0 & 9500.0 & 4.3 & 4.1 & 4.01 & 260.0 & 249.0 & 285.0 & 0.3 & -- & -- & -20.1&\\
HD73450 & 7200.0 & 7270.0 & 7850.0 & 3.1 & 4.2 & 4.2 & 120.0 & 125.0 & 135.0 & -0.4 & -0.03 & -0.03 & 13.7&V\\
HD73574 & 7300.0 & 7662.0 & 7836.0 & 3.1 & 4.0 & 4.0 & 90.0 & 89.9 & 97.45 & -0.4 & 0.1 & 0.1 & 26.3&\\
HD73763 & 7400.0 & 7850.0 & 7927.0 & 3.4 & -- & -- & 130.0 & 115.0 & 159.0 & -0.3 & -- & -- & 29.8&V\\
HD74028 & 7400.0 & 7750.0 & 7850.0 & 3.2 & 4.5 & 4.5 & 130.0 & 141.0 & 165.0 & -0.4 & 0.09 & 0.09 & 10.3&V\\
HD74050 & 7400.0 & 7400.0 & 7600.0 & 3.3 & 3.66 & 3.93 & 150.0 & 145.0 & 145.0 & -0.6 & 0.15 & 0.15 & 3.4&V\\
HD74135 & 6800.0 & 6800.0 & 7400.0 & 2.6 & 3.2 & 3.6 & 90.0 & -- & -- & -0.4 & 0.14 & 0.14 & 26.7&\\
HD74587 & 7100.0 & 7335.0 & 7671.5 & 2.7 & 4.2 & 4.2 & 90.0 & -- & -- & -0.1 & 0.31 & 0.31 & 27.0&V\\
HD74656 & 7400.0 & 7400.0 & 8000.0 & 3.0 & 3.99 & 3.99 & 25.0 & -- & -- & 0.2 & 0.48 & 0.48 & 31.1&\\
HD74718 & 7200.0 & 7443.0 & 7469.0 & 3.1 & -- & -- & 130.0 & -- & -- & -0.5 & -- & -- & 6.8&\\
HD75137 & 10700.0 & 10142.0 & 9777.5 & 4.2 & 4.1 & 4.1 & 140.0 & 135.0 & 128.0 & 0.1 & -- & -- & 3.4&\\
HD76398 & 7800.0 & 7650.0 & 7650.0 & 3.3 & 4.1 & 4.15 & 110.0 & 130.0 & 133.0 & -0.2 & -- & -- & -3.1&\\
HD76644 & 7600.0 & 7703.0 & 7769.0 & 3.5 & 4.19 & 4.19 & 120.0 & 140.0 & 150.0 & -0.3 & -0.03 & -0.03 & -3.4&\\
HD79439 & 7500.0 & 7547.0 & 8056.5 & 3.5 & 3.98 & 4.2 & 150.0 & 155.0 & 156.0 & -0.5 & -0.04 & -0.04 & -23.0&V\\
HD88960 & 11000.0 & 10377.0 & 9405.0 & 4.4 & 4.14 & 4.12 & 250.0 & 254.0 & 233.0 & 0.1 & -- & -- & -17.1&\\
HD90470 & 8000.0 & 8138.0 & 8247.5 & 3.6 & 4.2 & 4.2 & 110.0 & 105.0 & 90.0 & -0.3 & -- & -- & -6.6&\\
HD91312 & 7400.0 & 7600.0 & 7828.0 & 3.3 & 4.18 & 4.19 & 115.0 & 123.0 & 128.0 & -0.4 & -- & -- & 3.4&SB\\
HD92769 & 7600.0 & 7601.0 & 7601.0 & 3.7 & 4.1 & 4.2 & 190.0 & 202.0 & 212.0 & -0.6 & -0.2 & -0.15 & 0.0&\\
HD95418 & 9300.0 & 9342.0 & 9470.0 & 3.7 & 3.82 & 3.96 & 50.0 & 48.0 & 41.0 & 0.2 & 0.22 & 0.01 & -13.4&V\\
HD95804 & 7100.0 & 7303.0 & 7600.0 & 3.4 & 3.3 & 3.82 & 130.0 & -- & -- & -0.5 & -- & -- & -3.4&\\
HD97633 & 9000.0 & 9000.0 & 9302.0 & 3.4 & 3.45 & 3.6 & 25.0 & 24.0 & 22.0 & 0.0 & 0.03 & 0.04 & 3.4&V\\
HD100215 & 6900.0 & 7000.0 & 7000.0 & 3.4 & 4.15 & 4.15 & 20.0 & -- & -- & -0.3 & -0.04 & 0.05 & -23.2&V\\
HD106591 & 8100.0 & 8564.0 & 8600.0 & 3.6 & 4.12 & 4.16 & 200.0 & 210.0 & 210.0 & -0.6 & -0.35 & -0.35 & -21.7&V\\
HD111270 & 7500.0 & 7350.0 & 7870.0 & 3.2 & 4.18 & 4.24 & 90.0 & 100.0 & 105.0 & -0.3 & -- & -- & -12.9&\\
HD116656 & 9700.0 & 9676.0 & 9135.0 & 4.0 & -- & -- & 80.0 & 61.0 & 32.0 & 0.2 & 0.0 & -0.03 & -50.8&SB\\
HD118232 & 7700.0 & 8180.0 & 8489.0 & 3.4 & 3.79 & 3.99 & 140.0 & 159.0 & 162.0 & -0.4 & -- & -- & -24.4&\\
HD120047 & 7400.0 & 7593.0 & 7987.0 & 3.4 & 4.2 & 4.24 & 200.0 & 205.0 & 238.0 & -0.5 & -- & -- & -17.1&\\
HD121164 & 7600.0 & 7600.0 & 7918.0 & 3.4 & 4.18 & 4.19 & 60.0 & 60.0 & 75.0 & -0.1 & -- & -- & -17.0&\\
HD124675 & 7400.0 & 7500.0 & 7111.0 & 3.1 & 3.48 & 4.25 & 110.0 & 115.0 & 77.5 & -0.4 & -0.25 & -0.11 & -19.8&V\\
HD124713 & 7500.0 & 7600.0 & 7850.0 & 3.5 & 4.2 & 4.26 & 70.0 & 75.0 & 84.0 & -0.3 & -- & -- & -3.4&\\
HD129002 & 9400.0 & 9500.0 & 9230.0 & 4.0 & 4.1 & 4.18 & 90.0 & 86.0 & 86.0 & 0.1 & -- & -- & -23.1&\\
HD130109 & 11000.0 & 10110.0 & 9522.0 & 4.2 & 4.1 & 4.09 & 300.0 & 305.0 & 308.25 & -0.2 & -0.41 & -0.41 & -32.5&V\\
HD132145 & 10000.0 & 9500.0 & 9230.0 & 4.2 & 4.1 & 4.1 & 14.0 & 15.0 & 15.0 & 0.0 & -- & -- & -13.4&\\
HD133962 & 10500.0 & 9751.0 & 9230.0 & 4.3 & 4.17 & 4.14 & 60.0 & 55.0 & 49.0 & 0.1 & -- & -- & -13.4&V\\
HD139006 & 10900.0 & 10342.0 & 9825.0 & 4.3 & 3.82 & 3.82 & 120.0 & 118.5 & 132.5 & 0.2 & -- & -- & -30.7&EB\\
HD140159 & 8100.0 & 8302.0 & 9194.5 & 3.7 & 3.86 & 3.86 & 130.0 & 120.0 & 120.0 & -0.8 & -- & -- & -20.4&\\
HD142500 & 7500.0 & 7600.0 & 8088.0 & 3.5 & 4.2 & 4.2 & 210.0 & 210.0 & 228.0 & -0.6 & -- & -- & -50.2&V\\
HD143894 & 8300.0 & 8400.0 & 8720.0 & 3.6 & 4.04 & 4.11 & 130.0 & 128.0 & 120.0 & -0.3 & 0.38 & 0.38 & -34.0&\\
HD145454 & 10900.0 & 11765.0 & 9790.0 & 4.2 & 4.13 & 4.11 & 300.0 & 280.0 & 275.0 & -0.3 & -- & -- & -40.2&\\
HD154431 & 7500.0 & 7800.0 & 8140.5 & 3.5 & 4.2 & 4.25 & 100.0 & 120.0 & 123.0 & -0.5 & -0.11 & -0.11 & -27.0&\\
HD156653 & 9100.0 & 8978.0 & 9230.0 & 3.5 & 4.1 & 4.1 & 40.0 & 43.0 & 43.0 & -0.1 & -- & -- & 3.3&\\
HD157087 & 8400.0 & 8600.0 & 8770.0 & 3.1 & 3.4 & 3.4 & 12.0 & 12.2 & 15.0 & -0.1 & -- & -- & -10.1&\\
HD157778 & 10000.0 & 9520.0 & 8932.5 & 3.6 & -- & -- & 80.0 & 81.0 & 195.0 & 0.0 & -- & -- & -26.7&\\
HD162579 & 9800.0 & 9785.0 & 9000.0 & 4.1 & 4.02 & 4.11 & 120.0 & 122.0 & 123.0 & 0.5 & -- & -- & -64.3&\\
HD168913 & 7600.0 & 7461.0 & 7628.0 & 4.4 & 4.08 & 4.08 & 18.0 & 20.0 & 22.7 & -0.2 & -- & -- & 16.7&SB\\
HD169959 & 7300.0 & 7330.0 & 9698.0 & 2.6 & 3.93 & 4.03 & 140.0 & 70.0 & 41.0 & -1.4 & -0.1 & -0.05 & -49.5&BS\\
HD170973 & 10400.0 & 10309.0 & 10529.5 & 3.3 & 3.58 & 3.83 & 10.0 & 12.8 & 18.0 & 0.9 & 0.75 & 0.74 & -15.3&V\\
HD171586 & 8600.0 & 8607.0 & 8607.0 & 3.3 & 3.42 & 3.7 & 50.0 & -- & -- & 0.6 & 0.47 & -0.1 & 3.3&V\\
HD172167 & 9500.0 & 9511.0 & 9520.0 & 4.0 & 4.0 & 3.98 & 25.0 & 25.0 & 23.0 & -0.5 & -0.5 & -0.53 & -16.4&\\
HD173654 & 8200.0 & 8085.0 & 8834.0 & 3.6 & 4.15 & 4.15 & 60.0 & 61.0 & 61.0 & 0.5 & -- & -- & 16.8&SB\\
HD174567 & 9900.0 & 9790.0 & 10200.0 & 3.5 & 3.55 & 3.55 & 12.0 & 13.3 & 15.0 & -0.2 & -0.07 & 0.15 & -13.2&\\
HD184006 & 7500.0 & 7299.0 & 8200.0 & 2.9 & 3.77 & 3.98 & 220.0 & 220.0 & 226.0 & -0.5 & -- & -- & -33.0&\\
HD186689 & 7700.0 & 7906.0 & 8061.5 & 3.7 & 4.21 & 4.24 & 30.0 & 31.0 & 42.0 & -0.3 & -0.05 & -0.05 & -30.1&\\
HD193369 & 10100.0 & 9793.0 & 9358.0 & 4.4 & 4.3 & 4.25 & 110.0 & 110.0 & 102.0 & 0.4 & -- & -- & -16.7&\\
HD197950 & 7400.0 & 7500.0 & 7850.0 & 3.3 & 4.3 & 4.3 & 150.0 & 166.0 & 175.0 & -0.5 & -- & -- & -32.8&\\
HD198552 & 9000.0 & 9001.0 & 9292.0 & 3.8 & 4.1 & 4.18 & 50.0 & 50.0 & 52.0 & -0.1 & -- & -- & 3.4&\\
HD199095 & 10400.0 & 10500.0 & 9176.0 & 4.2 & -- & -- & 30.0 & 31.3 & 32.0 & 0.0 & -- & -- & 0.0&V\\
HD199254 & 7900.0 & 8146.0 & 8456.0 & 3.5 & 4.19 & 4.19 & 140.0 & 145.0 & 153.5 & -0.4 & -- & -- & -6.8&\\
HD203280 & 7300.0 & 7600.0 & 7773.0 & 3.2 & 4.2 & 4.2 & 190.0 & 196.0 & 206.0 & -0.4 & -- & -- & -22.6&V\\
HD204414 & 9400.0 & 9477.0 & 9477.0 & 4.0 & 4.1 & 4.19 & 80.0 & 79.0 & 81.0 & 0.4 & -- & -- & -9.9&\\
HD206677 & 7300.0 & 7600.0 & 7762.0 & 3.2 & 4.2 & 4.24 & 100.0 & 115.0 & 123.0 & -0.5 & -- & -- & -3.1&\\
HD209625 & 7400.0 & 7613.0 & 7858.0 & 2.6 & 3.73 & 3.73 & 12.0 & 14.2 & 26.5 & 0.2 & 0.27 & 0.3 & 16.4&SB\\
HD210418 & 8000.0 & 7951.0 & 8570.0 & 3.7 & 4.0 & 4.02 & 150.0 & 144.0 & 136.0 & -0.8 & -0.38 & -0.38 & -18.6&V\\
HD210715 & 7800.0 & 8180.0 & 8276.5 & 3.7 & 4.17 & 4.18 & 130.0 & 132.0 & 144.0 & -0.4 & -- & -- & -20.1&\\
HD213320 & 10700.0 & 10869.0 & 10112.5 & 4.1 & 4.1 & 4.07 & 25.0 & 25.0 & 21.0 & 0.1 & 0.38 & 0.44 & 12.2&\\
HD214454 & 7100.0 & 7150.0 & 7261.0 & 2.8 & 3.68 & 3.75 & 90.0 & 90.0 & 100.2 & -0.5 & -0.2 & -0.2 & 3.4&\\
HD214994 & 9500.0 & 9500.0 & 9608.0 & 3.6 & 3.62 & 3.65 & 8.0 & 8.7 & 12.0 & 0.1 & 0.12 & 0.1 & 6.5&\\
HD217186 & 8400.0 & 8310.0 & 9190.0 & 3.4 & 4.1 & 4.16 & 70.0 & 60.0 & 57.5 & -0.4 & -- & -- & 0.0&\\
HD218396 & 7000.0 & 7033.0 & 7459.5 & 3.8 & 4.35 & 4.35 & 35.0 & 47.2 & 49.0 & -0.7 & -0.47 & -0.47 & -12.2&V\\
HD219485 & 10000.0 & 9790.0 & 9568.0 & 3.9 & 4.1 & 4.1 & 25.0 & 23.0 & 23.0 & -0.1 & -- & -- & -6.6&\\
HD220061 & 7400.0 & 7655.0 & 8065.0 & 3.4 & 3.78 & 3.85 & 130.0 & 135.0 & 146.0 & -0.7 & -0.28 & -0.28 & -3.4&V\\
HD220974 & 7700.0 & 7750.0 & 8016.0 & 3.5 & 4.14 & 4.17 & 90.0 & 101.0 & 102.0 & -0.3 & -- & -- & 0.0&\\
HD221774 & 7500.0 & 7737.0 & 8087.0 & 3.5 & -- & -- & 50.0 & -- & -- & -0.4 & -- & -- & 26.5&\\
HD221866 & 7300.0 & 7268.0 & 7908.0 & 3.3 & -- & -- & 35.0 & -- & -- & -0.1 & -- & -- & -9.8&V\\
HD222345 & 7300.0 & 7344.0 & 7667.0 & 3.3 & 4.1 & 4.15 & 80.0 & 97.0 & 102.0 & -0.2 & -- & -- & -6.8&\\
HD222603 & 7500.0 & 7450.0 & 7766.0 & 3.5 & 3.9 & 4.08 & 50.0 & 62.8 & 65.0 & -0.3 & -0.06 & 0.02 & 6.7&\\
HD223855 & 9900.0 & 9547.0 & 9520.0 & 4.0 & 4.1 & 4.1 & 70.0 & 60.0 & 54.85 & -0.1 & -- & -- & 0.0&\\
HD224624 & 7800.0 & 8164.0 & 8325.0 & 3.8 & -- & -- & 160.0 & -- & -- & -0.7 & -- & -- & -15.5&\\
HIP728 & 7400.0 & 7535.0 & 8019.0 & 3.1 & -- & -- & 40.0 & 55.6 & 60.0 & 0.0 & -- & -- & 3.1&SB\\
HIP24951 & 6800.0 & 7693.0 & 7693.0 & 4.1 & -- & -- & 25.0 & 23.0 & 19.8 & 0.3 & -- & -- & 13.4&SB\\
HIP56429 & 7600.0 & 7740.0 & 7950.0 & 3.3 & -- & -- & 8.0 & 10.0 & 12.4 & -0.4 & -- & -- & -13.3&SB\\
HIP058159 & 7800.0 & 7719.0 & 8542.0 & 3.2 & 3.0 & 3.45 & 8.0 & 10.0 & 16.0 & -0.8 & -0.19 & 0.18 & -55.7&SB\\
HIP60490 & 7300.0 & 7257.0 & 7488.0 & 3.6 & 4.16 & 4.16 & 110.0 & 128.0 & 150.0 & -0.3 & -- & -- & 3.0&SB\\
HIP109119 & 8000.0 & 8090.0 & 8090.0 & 3.3 & -- & -- & 20.0 & -- & -- & 0.2 & -- & -- & 3.3&\\
HR553 & 8000.0 & 8061.0 & 8400.0 & 3.6 & 3.73 & 3.85 & 70.0 & 70.0 & 73.0 & 0.0 & 0.05 & 0.07 & -6.6&SB\\
HR7596 & 10500.0 & 10115.0 & 8210.0 & 3.8 & 3.4 & 3.4 & 110.0 & 117.0 & 117.0 & 0.0 & 0.0 & -0.23 & -53.1&\\
omi Leo & 6800.0 & 6875.0 & 6690.0 & 4.0 & 3.42 & 3.23 & 16.0 & 15.0 & 15.0 & 0.3 & 0.13 & 0.63 & -26.6&SB\\ \hline
\end{supertabular}
\end{center}


\begin{thebibliography}{}
\bibitem[Bailer-Jones et al.(1998)]{bailer} Bailer-Jones, C.~A.~L., Irwin, M., \& von Hippel, T.\ 1998, \mnras, 298, 361 

\bibitem[Balega et al.(2012)]{balega12} Balega, Y.~Y., Dyachenko, V.~V., Maksimov, A.~F., et al.\ 2012, Astrophysical Bulletin, 
67, 44 
\bibitem[Baranne et al.(1996)]{baranne96} Baranne, A., Queloz, D., Mayor, M., et al.\ 1996, \aaps, 119, 373 

\bibitem[Blanco-Cuaresma et al.(2014)]{fgkstars} Blanco-Cuaresma, S., Soubiran, C., Jofr{\'e}, P., \& Heiter, U.\ 2014, \aap, 566, A98 

\bibitem[Borne(2013)]{borne13} Borne, K.\ 2013, Planets, Stars and Stellar Systems.~Volume 2: Astronomical Techniques, Software and Data, 
403 





\bibitem[Castelli \& Kurucz(2003)]{castelli03} Castelli, F., \& Kurucz, R.~L.\ 2003, Modelling of Stellar Atmospheres, 210, 20P 

\bibitem[Catanzaro 
\& Balona(2012)]{balona12} Catanzaro, G., \& Balona, L.~A.\ 2012, \mnras, 421, 1222 


\bibitem[Code et al.(1976)]{code76} Code, A.~D., Bless, R.~C., 
Davis, J., \& Brown, R.~H.\ 1976, \apj, 203, 417 


\bibitem[Cowley et al.(1969)]{cowley69} Cowley, A., Cowley, C., Jaschek, M., \& Jaschek, C.\ 1969, \aj, 74, 375 

\bibitem[Demmel (1997)]{demel97} Demmel, James W.\ 1997. Applied Numerical Linear Algebra. Soc. for Industrial and Applied Math., Philadelphia, PA, USA.

\bibitem[Eggleton \& Tokovinin(2008)]{Eggleton08} Eggleton, P.~P., \& Tokovinin, A.~A.\ 2008, \mnras, 389, 869 

\bibitem[Farah et al.(2015)]{farah15} Farah, W., Gebran, M., Paletou, F., \& Blomme, R.\ 2015, arXiv:1508.03978. Proceedings of the SF2A 2015 (http://sf2a.eu/spip/spip.php?article638) 



\bibitem[Gazzano et al.(2010)]{Gazzano} Gazzano, J.-C., de Laverny, P., Deleuil, M., et al.\ 2010, \aap, 523, A91 

\bibitem[Gebran et al.(2008)]{gebran08a} Gebran, M., Monier, R., \& Richard, O.\ 2008, \aap, 479, 189 


\bibitem[Gebran \& Monier(2008)]{gebran08b} Gebran, M., \& Monier, R.\ 2008, \aap, 483, 567 


\bibitem[Gebran et al.(2010)]{gebran10} Gebran, M., Vick, M., Monier, R., \& Fossati, L.\ 2010, \aap, 523, A71 

\bibitem[Gebran et al.(2014)]{gebranxi} Gebran, M., Monier, R., Royer, F., Lobel, A., \& Blomme, R.\ 2014, Putting A Stars into Context: Evolution, Environment, and Related Stars, 193 

\bibitem[Gilmore et al.(2012)]{ges} Gilmore, G., Randich, S., Asplund, M., et al.\ 2012, The Messenger, 147, 25 


\bibitem[Gray(1992)]{gray92} Gray, D.~F.\ 1992, Camb.~Astrophys.~Ser., Vol.~20,,  

\bibitem[Gray \& Garrison(1989)]{gray89} Gray, R.~O., \& Garrison, R.~F.\ 1989, \apjs, 70, 623 

\bibitem[Grevesse \& Sauval(1998)]{Grevesse98} Grevesse, N., \& Sauval, A.~J.\ 1998, Solar Composition and Its Evolution -- From Core to Corona, 161 


\bibitem[Heiter et al.(2002)]{heiter2002} Heiter, U., Kupka, F., van't Veer-Menneret, C., et al.\ 2002, \aap, 392, 619 

\bibitem[Hill et al.(2010)]{hill10} Hill, G., Gulliver, A.~F., \& Adelman, S.~J.\ 2010, \apj, 712, 250 

\bibitem[Hill \& Landstreet(1993)]{hill93} Hill, G.~M., \& Landstreet, J.~D.\ 1993, \aap, 276, 142 


    
\bibitem[{{Hubeny} \& {Lanz}(1992)}]{Hubeny} Hubeny, I. \& Lanz, T. 1992, A\&A, 262,501

\bibitem[Jolliffe(1986)]{Jolliffe} Jolliffe, I.~T.\ 1986, Springer Series in Statistics, Berlin: Springer, 1986,  

\bibitem[Kantor et al.(2007)]{kantor07} Kantor, J., Axelrod, T., 
Becla, J., et al.\ 2007, Astronomical Data Analysis Software and Systems 
XVI, 376, 3 




 \bibitem[Kudryavtsev et al.(2006)]{Kudryavtsev} Kudryavtsev, D.~O., Romanyuk, I.~I., Elkin, V.~G., \& Paunzen, E.\ 2006, \mnras, 372, 1804 
   
 \bibitem[{{Kurucz}(1992)}]{Kurucz} Kurucz, R.L. 1992, Rev. Mex. Astron. Astrofis., 23, 45

\bibitem[Kurucz(2005)]{kurucz05} Kurucz, R.~L.\ 2005, Memorie 
della Societa Astronomica Italiana Supplementi, 8, 14 



\bibitem[Landstreet et al.(2008)]{landstreet08} Landstreet, J.~D., Silaj, J., Andretta, V., et al.\ 2008, \aap, 481, 465 

\bibitem[Landstreet(2011)]{landstreet11} Landstreet, J.~D.\ 2011, \aap, 528, A132 


\bibitem[Lef{\`e}vre et al.(2009)]{lefevre09} Lef{\`e}vre, L., Michel, E., Aerts, C., et al.\ 2009, Communications in Asteroseismology, 158, 189 

\bibitem[Liu et al.(1991)]{liu91} Liu, T., Janes, K.~A., \& Bania, T.~M.\ 1991, \apj, 377, 141 

\bibitem[Malkov et al.(2012)]{malkov12} Malkov, O.~Y., Tamazian, V.~S., Docobo, J.~A., \& Chulkov, D.~A.\ 2012, \aap, 546, A69 


\bibitem[McCuskey(1956)]{McCuskey56} McCuskey, S.~W.\ 1956, \apjs, 2, 271 

\bibitem[Moultaka et al.(2004)]{moultaka} Moultaka, J., Ilovaisky, S.~A., Prugniel, P., \& Soubiran, C.\ 2004, \pasp, 116, 693 


\bibitem[Munari et al.(2005)]{rave2} Munari, U., Zwitter, T., \& Siebert, A.\ 2005, The Three-Dimensional Universe with Gaia, 576, 529 

\bibitem[Paletou et al.(2015)]{2015sf2a.conf...37P} Paletou, F., Glorian, 
J.-M., G{\'e}not, V., et al.\ 2015, SF2A-2015: Proceedings of the Annual 
meeting of the French Society of Astronomy and Astrophysics.~Eds.: 
F.~Martins, S.~Boissier, V.~Buat, L.~Cambr{\'e}sy, P.~Petit, pp.37-40, 37 



\bibitem[Paletou et al.(2015a)]{S4n} Paletou, F., B{\"o}hm, T., Watson, V., \& Trouilhet, J.-F.\ 2015, \aap, 573, A67 

\bibitem[Paletou et al.(2015b)]{dms} Paletou, F., Gebran, M., Houdebine, E.~R., \& Watson, V.\ 2015, \aap, 580, A78 

\bibitem[Paletou \& Zolotukhin(2014)]{2014arXiv1408.7026P} Paletou, F., \& Zolotukhin, I.\ 2014, arXiv:1408.7026 

\bibitem[Perryman et al.(2001)]{perryman01} Perryman, M.~A.~C., de Boer, K.~S., Gilmore, G., et al.\ 2001, \aap, 369, 339 

 
\bibitem[Petit et al.(2014)]{polar} Petit, P., Louge, T., Th{\'e}ado, S., et al.\ 2014, \pasp, 126, 469 




\bibitem[Przybilla \& Butler(2004)]{przybilla2004} Przybilla, N., \& Butler, K.\ 2004, \apj, 609, 1181 


\bibitem[Recio-Blanco et al.(2015)]{recioblanco15} Recio-Blanco, A., de Laverny, P., Allende Prieto, C., et al.\ 2015, arXiv:1510.00111 

\bibitem[Rees et al.(2000)]{rees00} Rees, D.~E., L{\'o}pez Ariste, A., Thatcher, J., \& Semel, M.\ 2000, \aap, 355, 759 



\bibitem[Re Fiorentin et 
al.(2007)]{fiorentin} Re Fiorentin, P., Bailer-Jones, C.~A.~L., Lee, Y.~S., et al.\ 2007, \aap, 467, 1373 


\bibitem[Renson 
\& Manfroid(2009)]{renson09} Renson, P., \& Manfroid, J.\ 2009, \aap, 498, 961 

\bibitem[Richard et al.(2002)]{richard02} Richard, O., Michaud, G.,\& Richer, J.\ 2002, IAU Colloq.~185: Radial and Nonradial Pulsationsn as Probes of Stellar Physics, 259, 270 

\bibitem[Richard et al.(2001)]{richard01} Richard, O., Michaud, 
G., \& Richer, J.\ 2001, \apj, 558, 377 

\bibitem[Richer et al.(2000)]{richer00} Richer, J., Michaud, G., \& Turcotte, S.\ 2000, \apj, 529, 338 


\bibitem[Roweis(1998)]{roweis} Roweis, S. 1998, in Advances in Neural Information Processing Systems (MIT Press), 626–632

\bibitem[Royer et 
al.(2002)]{royer02} Royer, F., Grenier, S., Baylac, M.-O., G{\'o}mez, A.~E., \& Zorec, J.\ 2002, \aap, 393, 897 


\bibitem[Royer et 
al.(2014)]{royer14} Royer, F., Gebran, M., Monier, R., et al.\ 2014, \aap, 562, A84 


\bibitem[Ryabchikova et al.(2015)]{ryabchikova2015} Ryabchikova, T., Piskunov, N., \& Shulyak, D.\ 2015, Physics and Evolution of Magnetic and Related Stars, 494, 308 

\bibitem[Smalley \& Dworetsky(1993)]{smalley93} Smalley, B., \& Dworetsky, M.~M.\ 1993, \aap, 271, 515 
 

\bibitem[Smalley(2004)]{2004IAUS..224..131S} Smalley, B.\ 2004, The A-Star Puzzle, 224, 131

\bibitem[Smalley(2005)]{smalley2005} Smalley, B.\ 2005, Memorie della Societa Astronomica Italiana Supplementi, 8, 130 


\bibitem[Steinmetz(2003)]{rave1} Steinmetz, M.\ 2003, GAIA Spectroscopy: Science and Technology, 298, 381


\bibitem[Takeda et al.(2008)]{takeda08} Takeda, Y., Kawanomoto, S., \& Ohishi, N.\ 2008, \apj, 678, 446 

\bibitem[Takeda et al.(2009)]{takeda09} Takeda, Y., Kang, D.-I., Han, I., Lee, B.-C., \& Kim, K.-M.\ 2009, \pasj, 61, 1165 

\bibitem[Tonry \& Davis(1979)]{ccf} Tonry, J., \& Davis, M.\ 1979, \aj, 84, 1511 

\bibitem[Vick et al.(2010)]{vick10} Vick, M., Michaud, G., Richer, J., \& Richard, O.\ 2010, \aap, 521, A62 

\bibitem[York et al.(2000)]{sdss} York, D.~G., Adelman, J., 
Anderson, J.~E., Jr., et al.\ 2000, \aj, 120, 1579 

\bibitem[Zahn(2005)]{zahn05} Zahn, J.-P.\ 2005, EAS 
Publications Series, 17, 157
\end{thebibliography}
\end{document}